\documentclass[aps,prb,reprint,groupedaddress,floatfix,showpacs]{revtex4-1}

\usepackage{graphicx}
\usepackage{dcolumn}
\usepackage{bm}
\usepackage{color}

\usepackage{ulem}
\begin{document}

\title{A first-principles study of carbon-related energy levels in GaN: Part II - Complexes formed by carbon and hydrogen, silicon or oxygen}

\author{Masahiko Matsubara}
\author{Enrico Bellotti}
\affiliation{Department of Electrical and Computer Engineering, Boston University, 8 St. Mary's Street, Boston, MA 02215, USA}


\date{\today}







\begin{abstract}
This work presents an in-depth investigation of the properties of complexes composed of
hydrogen, silicon or oxygen with carbon, which are major unintentional impurities
in undoped GaN.  This  manuscript is a complement to our previous work on carbon--carbon and carbon-vacancy
complexes.
We have employed a first-principles method using Heyd-Scuseria-Ernzerhof hybrid functionals
within the framework of generalized Kohn-Sham density functional theory.
Two H--C, four Si--C and five O--C complexes in different charge states have been considered.
After full geometry relaxations, formation energies, binding energies and both
thermal and optical transition levels were obtained.
The calculated energy levels have been systematically compared with the experimentally observed carbon
related trap levels. Furthermore, we computed vibrational frequencies for selected defect
complexes and defect concentrations were estimated in the low, mid and high
carbon doping scenarios considering two different cases where electrically
active defects: (a) only carbon and vacancies and (b) not only carbon
and vacancies but also hydrogen, silicon and oxygen. We confirmed that $\mathrm{C_N}$
is a dominant acceptor in GaN. In addition to it, substantial amount of
$\mathrm{Si_{Ga}-C_N}$ complex exists in a neutral form. This complex is a likely
candidate for unknown form of carbon observed in undoped $n$-type GaN.
\end{abstract}
    
\pacs{61.72.J-, 61.72.uj, 71.15.Mb, 71.55.Eq}

\maketitle

\section{Introduction\label{s:intro}}

Carbon is ubiquitous in GaN regardless of intentional doping or
growth techniques. The incorporated carbon introduce trap levels
within the band gap, which cause undesirable effect on the
performance of GaN-based opto- or power-electronics devices.
Therefore, it is important to identify the physical form of carbon,
which causes the trap levels, in order to control its impact of
device operation. Experimental methods such as deep level transient
spectroscopy (DLTS), deep level optical spectroscopy (DLOS) and
related techniques have been used for GaN samples with different
carbon concentration and multiple trap levels have been observed to
be function of the carbon presence
~\cite{armstrong05,shah12,polyakov13,honda12}. Unfortunately, these
techniques are not capable of identifying the original forms of the
trap levels directly. In this regard computational approach is of
great help to interpret the experimental results.

In a companion paper to this manuscript~\cite{MatsubaraC1} we have
focused on the complexes formed by pairs of two carbons
($\mathrm{C_{N}-C_{Ga}}$, $\mathrm{C_{I}-C_{N}}$ and
$\mathrm{C_{I}-C_{Ga}}$, where $\mathrm{C_{N}}$ is a C at N site,
$\mathrm{C_{Ga}}$ is a C at Ga site and $\mathrm{C_{I}}$ is an
interstitial C) and pairs of carbon and nearest neighbor vacancy
($\mathrm{C_{N}-V_{Ga}}$ and $\mathrm{C_{Ga}-V_{N}}$, where
$\mathrm{V_{Ga}}$ is a Ga vacancy and $\mathrm{V_{N}}$ is a N
vacancy), for which our knowledge is still limited, as candidates
for the origins of carbon related deep level traps observed in
experiments and tried to identify them by comparing our calculated
results with experimental data. As a result, we successfully
assigned carbon-related energies reported by different experimental
groups to our theoretically determined trap levels with specific
forms of C. However, the formation energies of some of these
complexes are substantial (e.g.\ the formation energy of the neutral
charge state of $\mathrm{C_N-V_{Ga}}$ complex is almost 10\,eV),
resulting in very low concentrations, much below the experimental
detection limit.
In the present article we extend our search of C-related complexes
to the ones formed by hydrogen, silicon or oxygen in order to
complement the results with the ones in our previous
work~\cite{MatsubaraC1}. These three elements are also well-known as
common unintentional impurities in GaN, alongside with carbon.

Silicon is used to obtain controlled $n$-type conductivity in
GaN~\cite{koide91,nakamura92si,rowland95,gotz96}. Oxygen was
experimentally identified as donor due to unintentional doping,
which leads to a very high $n$-type conductivity~\cite{wetzel97}.
Theoretical calculations have shown that the Si substituting Ga atom
($\mathrm{Si_{Ga}}$) and the O substituting nitrogen atom
($\mathrm{O_{N}}$) act as shallow donors with low formation
energies~\cite{neugebauer96,mattila97}. Complexes based on
$\mathrm{Si_{Ga}}$ and $\mathrm{O_N}$ have also been studied.
$\mathrm{O_N-C_N}$ has been indicated as a possible culprit for the
origins of yellow luminescence~\cite{demchenko13}. On the other
hand, a recent study concluded that neither $\mathrm{O_N-C_N}$ nor
$\mathrm{Si_{Ga}-C_N}$ can be the origins of the yellow luminescence in
GaN~\cite{christenson15}. It has also been speculated that hydrogen
forms complexes with impurities in GaN and could modify
its electronic behavior. Hydrogen atoms neutralize Mg acceptors in
GaN~\cite{nakamura92ptype,brandt94apl,brandt94prb} and complexes of multiple hydrogens with
native defects, in particular vacancies, have also been
investigated~\cite{vandewalle97,lyons15}. Furthermore, complexes of
hydrogen with carbon, have also been also studied in relation to the
blue luminescence~\cite{demchenko16}. As a result, it is possible
that complexes formed by these elements (Si, O and H) with carbon
lead to trap levels within the band gap, which are experimentally
detected as carbon related energy levels.

In this manuscript we provide a comprehensive study of C-based
complexes with other unintentional dopants (H, Si and O) within the
framework of generalized Kohn-Sham density functional theory using
Heyd-Scuseria-Ernzerhof hybrid density functionals. Formation
energies, binding energies and both thermal and optical activation
energies are obtained from the total energies of fully optimized
complex structures in the different charge states.
By assigning our calculated energy levels to the experimentally
measured ones we try to identify the origins of C-related trap
levels, especially for those whose exact physical forms are not
known yet. In addition, concentrations of C impurities/complexes are
estimated for each form of C impurities. Acceptor form of C
($\mathrm{C_N^{1-}}$) is known to be dominant in
GaN~\cite{boguslawski96,wright02,lyons10,lyons14}.
However, secondary ion mass spectroscopy (SIMS) experiment reported in
Ref.~\onlinecite{tompkins11} for $n$-type GaN indicated that carbon concentration
is always higher than the carrier concentration. This result implies that not all
of the carbon takes the form of $\mathrm{C_N^{1-}}$ and that substantial amount
of other forms of carbon would exist in $n$-type GaN.
Our concentration
analysis provides an potential candidate for this unknown form of
carbon.

This manuscripts is organized as follows: in
Section~\ref{s:methods}, the computational methods are briefly
outlined as the detail are similar to the ones presented in our
companion work~\cite{MatsubaraC1}. In Section~\ref{s:results}, we
will present our calculated results for H, Si and O impurities and
complexes with C. Our theoretical trap levels are assigned to
experimentally observed ones and concentration analysis is performed
in Section~\ref{s:discussion}. Section~\ref{s:conclusion} concludes
this paper.

\section{Methods\label{s:methods}}

Details of the computational methods were already given in the first
part of this work~\cite{MatsubaraC1}, and they are only briefly
outlined here. Our calculations were performed using
Heyd-Scuseria-Ernzerhof (HSE) hybrid density
functional~\cite{heyd03,heyd06} implemented in the VASP
code~\cite{kresse96,kresse99}. The mixing amount of exact exchange
was taken to be 28\%, which gives 3.45\,eV band gap value for bulk
GaN\@.

The formation energies of a defect ($D$) with a charge state $q$, $E^q_f\left(D,E_F\right)$,
are calculated from the total energy of the system, $E_{\mathrm{tot}}^q\left(D\right)$,
and its bulk counterpart, $E_{\mathrm{bulk}}$, as well as chemical potentials for the
defect type X, $\mu_{\mathrm{X}}$, as a function of Fermi energy ($E_F$), which is set to
zero at the valence band maximum ($E_v$), i.e.\
\begin{eqnarray}
E_{f}^{q}\left(D,E_{F}\right) & = & E_{\mathrm{tot}}^{q}\left(D\right)-E_{\mathrm{bulk}}-\sum_{\mathrm{X}}n_{\mathrm{X}}\mu_{\mathrm{X}} \nonumber\\
	&+&q\left(E_{F}+E_{v}\right)+\Delta E_{\mathrm{corr}}^{q}.\label{eq:Ef}
\end{eqnarray}
Here $n_\mathrm{X}$ is the number of defect type X, which are added to or removed
from the system, and the last term corresponds to the correction for the charged defects
in the finite supercell~\cite{freysoldt09,freysoldt11,sxdefectalign}.
The chemical potentials for Ga, N and C are obtained from bulk $\alpha$-Ga,
nitrogen molecule and diamond, respectively, whereas
those for H, O and Si are obtained as
follows. The chemical potential for H ($\mu_{\mathrm{H}}$) is determined
using an isolated H$_{2}$ molecule as a reference. The chemical potentials
for oxygen ($\mu_{\mathrm{O}}$) and silicon ($\mu_{\mathrm{Si}}$)
depend on the growth conditions due to the possible formations of
$\mathrm{Ga_{2}O_{3}}$ and $\mathrm{Si_{3}N_{4}}$, respectively.
As a result, they must satisfy specific conditions shown below. For
$\mathrm{Si_{3}N_{4}}$, the condition is written as
\begin{eqnarray}
3\mu_{\mathrm{Si}}+4\mu_{\mathrm{N}} & = & 3E\mathrm{(Si)}+2E\mathrm{(N_{2})}+\Delta H_{f}\mathrm{(Si_{3}N_{4})},\label{eq:si3n4}
\end{eqnarray}
where $E\mathrm{(Si)}$, $E\mathrm{(N_{2})}$ and $\Delta H_{f}\mathrm{(Si_{3}N_{4})}$
are the energies of bulk silicon and nitrogen molecule and the formation
enthalpy of the $\beta$-$\mathrm{Si_{3}N_{4}}$, respectively. Our calculated
value for $\Delta H_{f}\mathrm{(Si_{3}N_{4})}$ is $-$8.52\,eV,
which is in good agreement with the experimental value ($-$8.58\,eV)~\cite{ohare99}.
With the Eq.~(\ref{eq:si3n4}) and the condition for GaN
\begin{eqnarray}
\mu_{\mathrm{Ga}}+\mu_{\mathrm{N}} & = & E\mathrm{(Ga)}+\frac{1}{2}E\mathrm{(N_{2})}+\Delta H_{f}\mathrm{(GaN)},\label{eq:gan}
\end{eqnarray}
in Ga-rich limit $\mu_{\mathrm{Si}}$ is computed as
\begin{eqnarray}
\mu_{\mathrm{Si}}\mathrm{(Ga-rich)} & = & E\mathrm{(Si)}+\frac{1}{3}\Delta H_{f}\mathrm{(Si_{3}N_{4})}-\frac{4}{3}\Delta H_{f}\mathrm{(GaN)},\label{eq:musiga} \nonumber\\
\end{eqnarray}
whereas in N-rich limit it is computed as
\begin{eqnarray}
\mu_{\mathrm{Si}}\mathrm{(N-rich)} & = & E\mathrm{(Si)}+\frac{1}{3}\Delta H_{f}\mathrm{(Si_{3}N_{4})}.\label{eq:musin}
\end{eqnarray}
For $\mathrm{Ga_{2}O_{3}}$, it is also written as
\begin{eqnarray}
2\mu_{\mathrm{Ga}}+3\mu_{\mathrm{O}} & = & 2E\mathrm{(Ga)}+\frac{3}{2}E\mathrm{(O_{2})}+\Delta H_{f}\mathrm{(Ga_{2}O_{3})},\label{eq:ga2o3}
\end{eqnarray}
where $E\mathrm{(Ga)}$, $E\mathrm{(O_{2})}$ and $\Delta H_{f}\mathrm{(Ga_{2}O_{3})}$
are the energies of bulk $\alpha$-Ga and oxygen molecule and the
formation enthalpy of the $\beta$-$\mathrm{Ga_{2}O_{3}}$, respectively.
Our calculated value for $\Delta H_{f}\mathrm{(Ga_{2}O_{3})}$ is
$-$10.00\,eV, which is close to the experimental value ($-$11.29\,eV)~\cite{lide05}.
With the Eqs.~(\ref{eq:gan}) and~(\ref{eq:ga2o3}), $\mu_{\mathrm{O}}$
in Ga-rich limit becomes
\begin{eqnarray}
\mu_{\mathrm{O}}\mathrm{(Ga-rich)} & = & \frac{1}{2}E\mathrm{(O_{2})}+\frac{1}{3}H_{f}\mathrm{(Ga_{2}O_{3})},\label{eq:muoga}
\end{eqnarray}
whereas in N-rich it becomes
\begin{eqnarray}
\mu_{\mathrm{O}}\mathrm{(N-rich)} & = & \frac{1}{2}E\mathrm{(O_{2})}+\frac{1}{3}H_{f}\mathrm{(Ga_{2}O_{3})}-\frac{2}{3}H_{f}\mathrm{(GaN)}.\label{eq:muon} \nonumber\\
\end{eqnarray}
Using the computed formation energies, the thermodynamic transition
level between charge states $q$ and $q^{\prime}$ is obtained with
\begin{eqnarray}
\epsilon\left(q/q^{\prime}\right) & = & \frac{E_{f}^{q}\left(D,E_{F}=0\right)-E_{f}^{q^{\prime}}\left(D,E_{F}=0\right)}{q^{\prime}-q},\label{eq:eqq}
\end{eqnarray}
which corresponds to the experimentally observed trap levels by thermal techniques
such as DLTS\@. By adding Franck-Condon shift, optical levels can be also obtained
from eq.~(\ref{eq:eqq}), which corresponds to the experimentally observed trap
levels by optical techniques such as DLOS\@.

The concentration, $\left[\mathrm{C}\right]$, of an impurity with the formation
energy $E_{f}$ is obtained with
\begin{eqnarray}
\left[\mathrm{C}\right] & = & N_{\mathrm{site}}N_{\mathrm{config}}\exp\left(\frac{-E_{f}}{k_{\mathrm{B}}T}\right), \label{eq:C}
\end{eqnarray}
where $N_{\mathrm{site}}$ is the number of sites per unit volume and
$N_{\mathrm{config}}$ is the number of equivalent configurations the
defect can take. For example, in the case of substitutional impurity such
as $\mathrm{C_N}$, $N_{\mathrm{config}}=1$ and in the case of $\mathrm{Si_{Ga}-C_{N}}$
complex, $N_{\mathrm{config}}=4$.

\section{Results\label{s:results}}

\subsection{Hydrogen Impurities\label{ss:hydrogen}}

In this section we discuss interstitial H and its complex with two
substitutional impurity of carbon, i.e.\ $\mathrm{C_{N}}$ and
$\mathrm{C_{Ga}}$. First we determine the most favorable
configurations of $\mathrm{H_{I}}$ in various charge states. Then we
consider their complexes with $\mathrm{C_{N}}$/$\mathrm{C_{Ga}}$ and
obtain the lowest energy states.

\subsubsection{Interstitial Hydrogen\label{sss:HI}}

As initial configurations for $\mathrm{H_{I}}$, we considered
following six different sites, i.e.\ $\mathrm{BC_{\parallel}}$,
$\mathrm{BC_{\perp}}$, $\mathrm{AB_{\parallel}}$,
$\mathrm{AB_{\perp}}$, octahedral and tetrahedral interstitial
positions. BC represents the bond-centered sites and AB represents
the anti-bonding sites. Both BC and AB sites have two different
variations. For example, in the case of the BC position, when H is
located between Ga--N bonds parallel to $c$-axis, it is denoted as
$\mathrm{BC_{\parallel}}$, whereas when H is located between Ga--N
bonds (virtually) perpendicular to $c$-axis, it is denoted as
$\mathrm{BC_{\perp}}$. Then all the configurations were fully
relaxed within HSE\@.

\begin{figure*}
\begin{centering}
\includegraphics[width=0.25\textwidth]{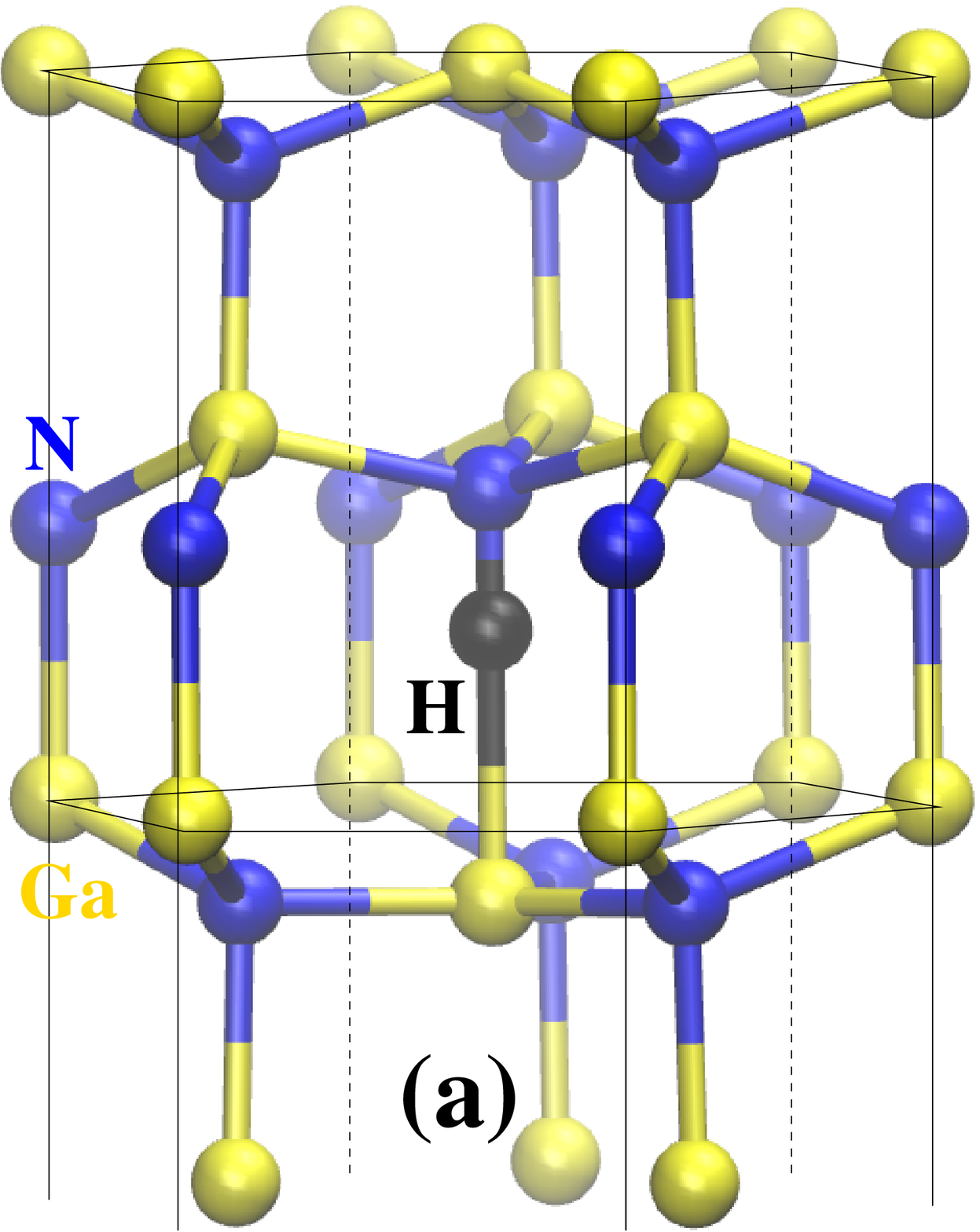}\includegraphics[width=0.25\textwidth]{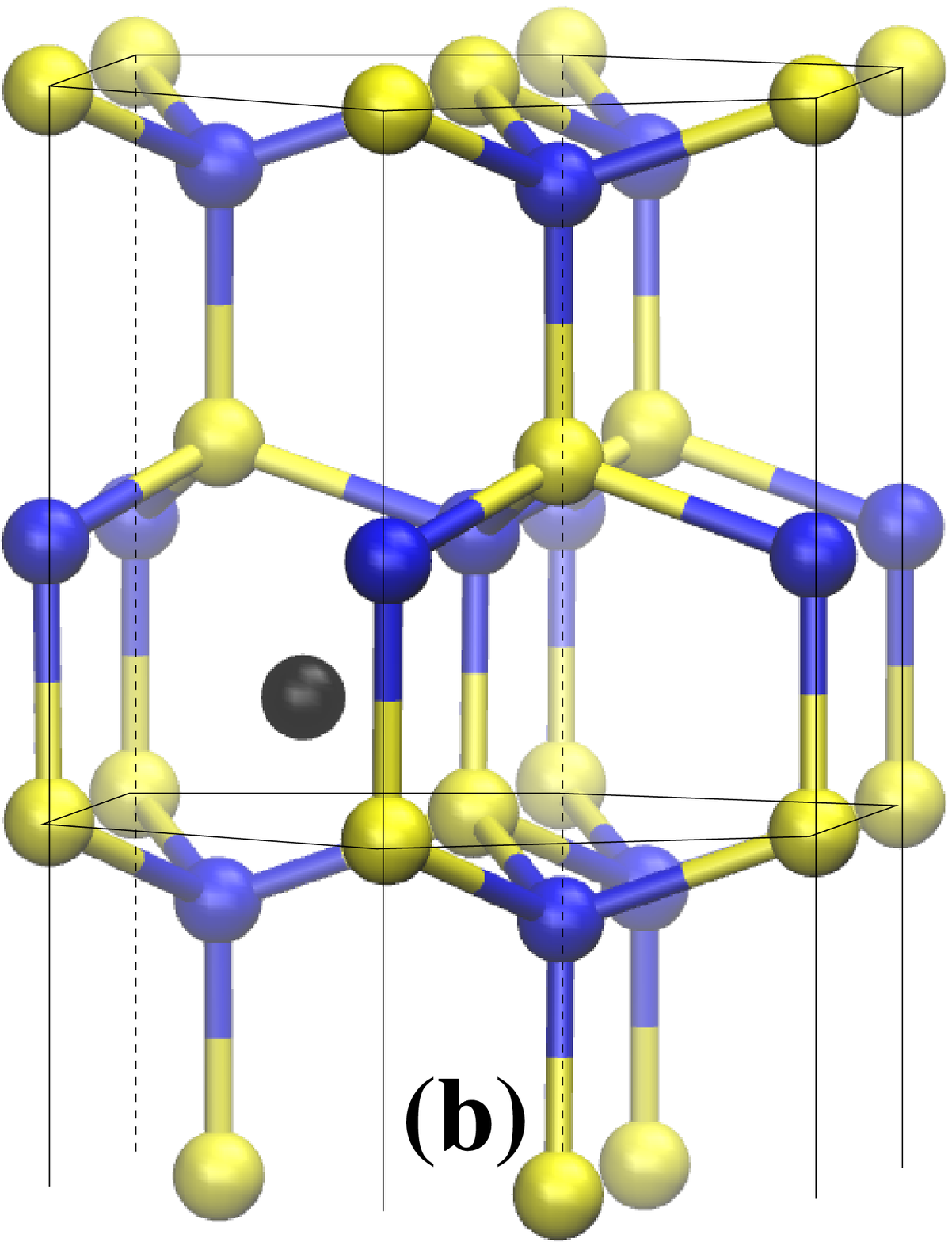}\includegraphics[width=0.25\textwidth]{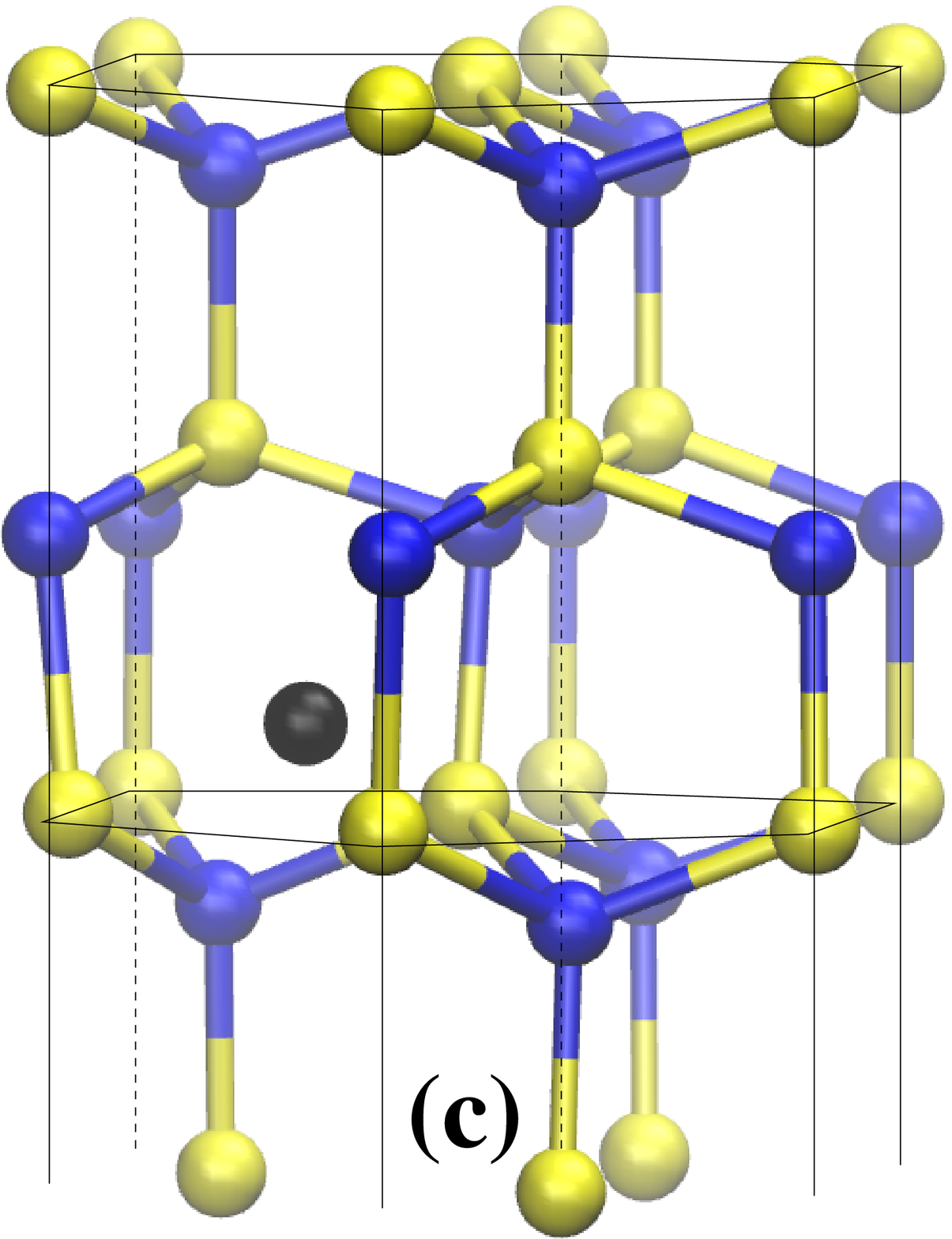}\includegraphics[width=0.25\textwidth]{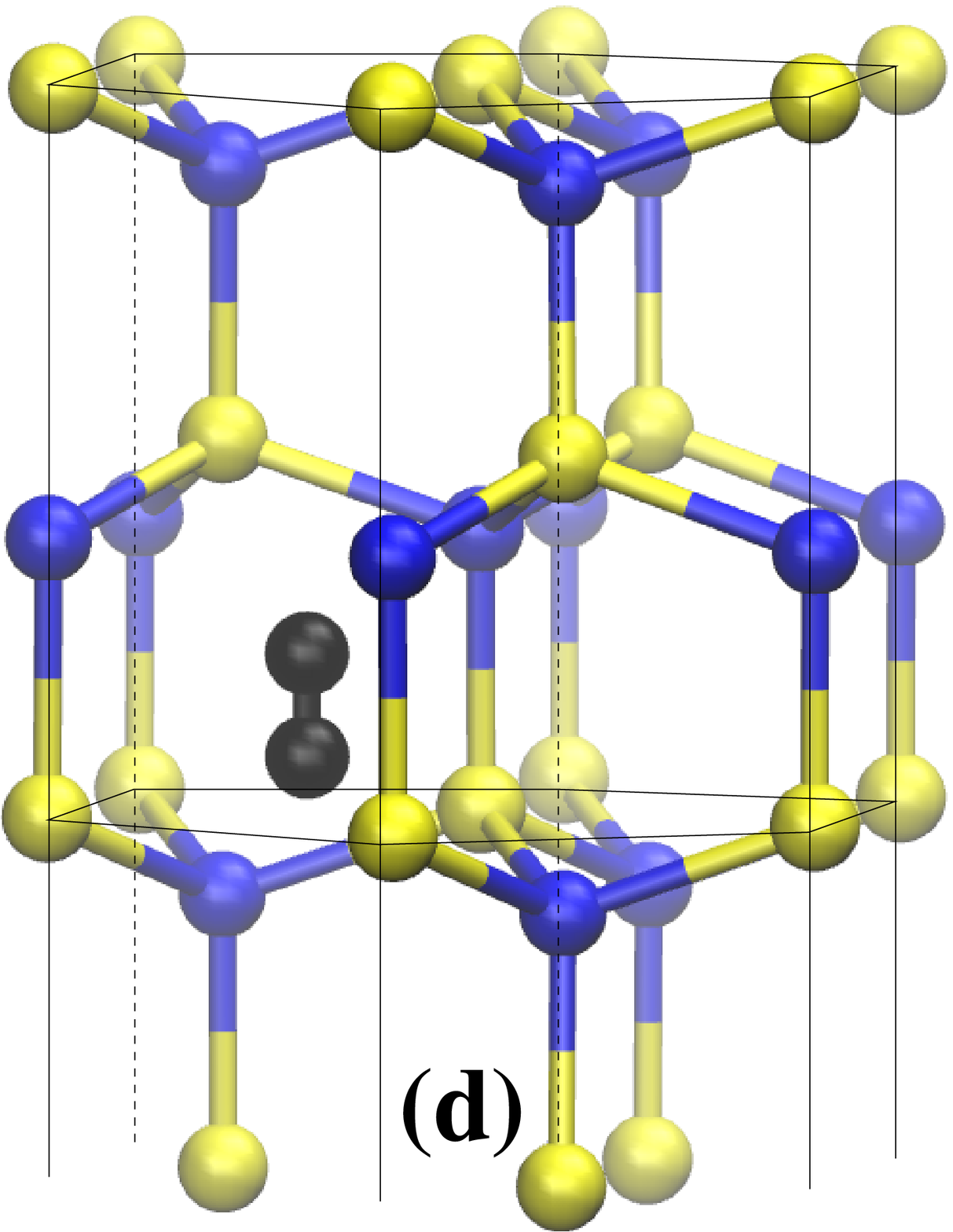}
\par\end{centering}
\caption{Stable positions of H in GaN after relaxation in HSE. H atoms are
denoted by black spheres, while Ga and N atoms are denoted by yellow
and blue spheres, respectively. (a) $\mathrm{BC_{\parallel}}$ position
in the 1+ charge state. (b) octahedral position in the 0 (neutral) charge state. (c)
octahedral position in the $1-$ charge state. Also Stable position of
$\mathrm{H_{2}}$ in the 0 charge state is provided in (d).\label{f:Hi}}
\end{figure*}

For the 1+ charge state, $\mathrm{BC_{\parallel}}$ position is the
most stable. This structure is given in FIG.~\ref{f:Hi} (a). The
distance between H and N ($d_{\mathrm{H-N}}$) is 1.01\,\AA, whereas
the distance between H and Ga ($d_{\mathrm{H-Ga}}$) is 1.93\,\AA.
The Ga atom is pushed down by the H atom and is located in the
N-atom plane.

For the 0 (neutral) and $1-$ charge states, H takes the octahedral
interstitial position. The structures are shown in FIGs.~\ref{f:Hi}
(b) and (c) for the 0 and $1-$ charge states, respectively. For the
0 charge state, average $d_{\mathrm{H-N}}$ is 2.28\,\AA, while
for the $1-$ charge state, average $d_{\mathrm{H-N}}$ is 2.33\,\AA.
These structural properties are very similar to those obtained in
previous LDA/GGA based calculations~\cite{myers00,wright03}.

\begin{figure}
\begin{centering}
\includegraphics[width=\columnwidth]{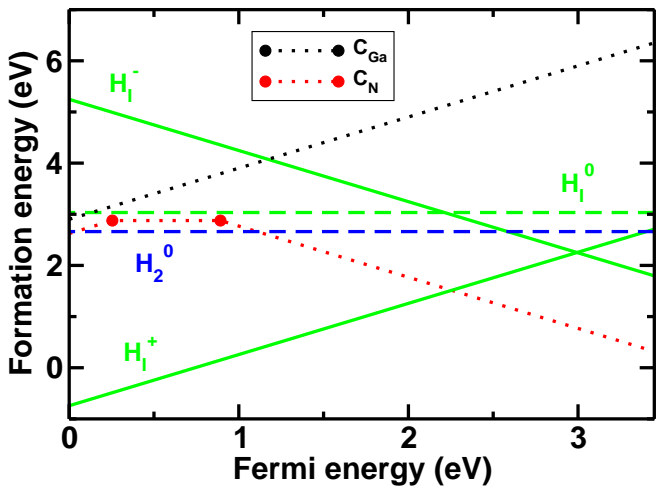}
\par\end{centering}
\caption{Formation energies of H in GaN as a function of Fermi energy. Formation
energies for 1+ and $1-$ charge states of $\mathrm{H_{I}}$ are shown
in green solid lines, whereas those for neutral $\mathrm{H_{I}}$
and $\mathrm{H_{2}}$ are shown in green and blue dashed lines, respectively.
The formation energies for $\mathrm{C_{Ga}}$ and $\mathrm{C_{N}}$ in
Ga-rich conditions are also given as references in black and red dotted lines,
respectively. \label{f:EfHi}}
\end{figure}

The calculated formation energies for $\mathrm{H_{I}}$ in different
charge states are shown in FIG.~\ref{f:EfHi}. The formation energies
for $\mathrm{C_{N}}$ both in Ga-rich and N-rich conditions are also
given as references with dotted lines in the figure.
$\mathrm{H_{I}^{+}}$ is the most stable state up to 2.99\,eV above
the VBM and $\mathrm{H_{I}}^{-}$ becomes the most stable when the
Fermi energy is higher than 2.99\,eV. The neutral
$\mathrm{H_{I}^{0}}$ never becomes a stable charge state at all
Fermi levels within the band gap.
The $\left(+/-\right)$ transition
level of 2.99\,eV in our HSE result is in good agreement with previously obtained
HSE results~\cite{lyons12,demchenko16}, although it is in contrast to the values obtained in
previous LDA~\cite{myers00} and GGA~\cite{wright03} calculations,
which were 2.10\,eV and 1.98\,eV, respectively.

We also considered a $\mathrm{H_{2}}$ molecule as a possible form of
the interstitial hydrogen in GaN\@. The octahedral and tetrahedral
positions were selected as initial positions of the $\mathrm{H_{2}}$
molecule and they were fully relaxed within HSE\@. We found that
only the 0 charge state is the most stable at all the Fermi energy
as shown in blue dashed line in FIG.~\ref{f:EfHi}. The structure is
shown in FIG.~\ref{f:Hi} (d). The direction of the H--H bond of
$\mathrm{H_{2}}$ molecule is parallel to the $c$-axis and the bond
length is 0.73\,\AA. $\mathrm{H_{2}^{0}}$ has lower formation energy
than $\mathrm{H_{I}^{0}}$, but never becomes stable because its
formation energy is always higher than those of $\mathrm{H_{I}^{+}}$
or $\mathrm{H_{I}^{-}}$. This result is in contrast to the previous
LDA/GGA based calculations, where $\mathrm{H_{2}^{0}}$ was expected
to be the most stable state around the middle of the band
gap~\cite{myers00,wright03}. The difference between the previous
results and our HSE result stem from the difference of the reference
energy for H atom rather than the difference introduced by the
functionals that were employed. In
Refs.~\onlinecite{myers00,wright03}, $\mu_{\mathrm{H}}$ is obtained
from the energy of a H atom, whereas we obtain it from the half of
the energy of a $\mathrm{H_{2}}$ molecule. Indeed, we found that if
we use the $\mu_{\mathrm{H}}$ taken from a H atom in our HSE
calculations, the formation energy of $\mathrm{H_{2}^{0}}$ is
lowered and becomes stable around the middle of the band gap as in
the cases of LDA/GGA\@.

\subsubsection{Complexes of Hydrogen with Carbon\label{sss:CH}}

We turn our attention to the complex made of C and H. Both $\mathrm{C_{N}}$
and $\mathrm{C_{Ga}}$ were considered as a constituent of the complex.
 One of the N (Ga) atoms is replaced by C to make a $\mathrm{C_{N}}$
($\mathrm{C_{Ga}}$) and a H atom is placed interstitial positions
around the $\mathrm{C_{N}}$ ($\mathrm{C_{Ga}}$) like in the case
of $\mathrm{H_{I}}$ and the structures are fully relaxed within HSE.

\begin{figure}
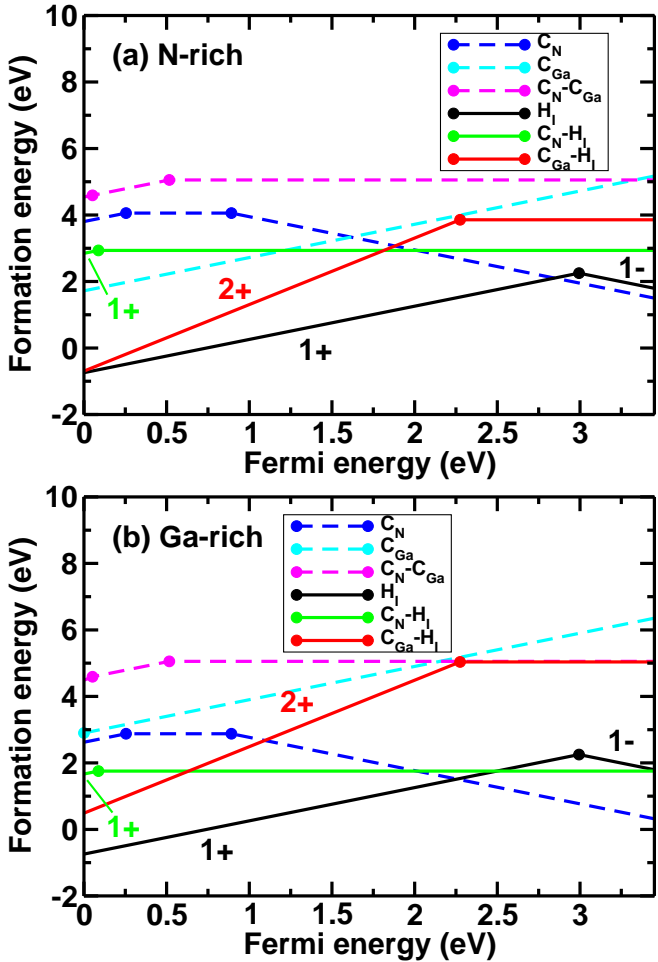

\begin{centering}
\includegraphics[width=\columnwidth]{fHiC-NrichwCGaCN}

\includegraphics[width=\columnwidth]{fHiC-GarichwCGaCN}
\par\end{centering}
\caption{Formation energies of $\mathrm{C_{N}-H_{I}}$ (green solid line) and
$\mathrm{C_{Ga}-H_{I}}$ (red solid line) complexes in GaN as a function
of Fermi energy (a) in N-rich and (b) in Ga-rich conditions. Formation
energies for $\mathrm{C_{N}}$ (blue dashed line), $\mathrm{C_{Ga}}$ (cyan dashed line), $\mathrm{C_{N}-C_{Ga}}$
(magenta dashed line) and $\mathrm{H_{I}}$ (black solid line) are
also given as references.\label{f:EfCNHi}}
\end{figure}

\begin{figure}
\begin{centering}
\includegraphics[width=\columnwidth]{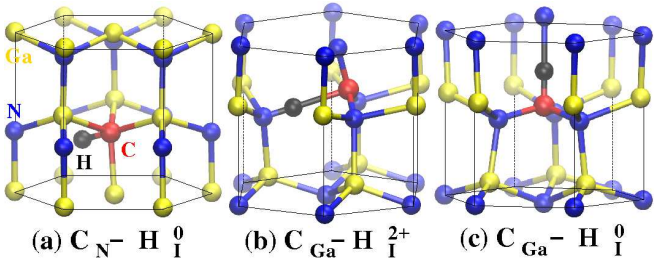}
\par\end{centering}
\caption{(a) Stable position of $\mathrm{C_{N}-H_{I}}$ complex in GaN after
the relaxation by HSE. H atom is denoted by black sphere, while C
atom is denoted by red sphere. $\mathrm{H_{I}}$ is located at the
$\mathrm{AB{}_{\perp}}$ position with 1.10\,\AA\ distance from
$\mathrm{C_{N}}$. Stable positions of $\mathrm{C_{Ga}-H_{I}}$ complex
in the (b) 2+ and (c) 0 charge states after the relaxation by HSE.
In the 2+ charge state, the complex is the most stable in the BC$_{\perp}$
position with $\mathrm{C-H}$ distance of 1.85\,\AA, while in the
0 charge state it is the most stable in the BC$_{\parallel}$ position
with $\mathrm{C-H}$ distance of 1.08\,\AA.\label{f:CNHi}}
\end{figure}

\begin{figure}
\begin{centering}
\includegraphics[width=\columnwidth]{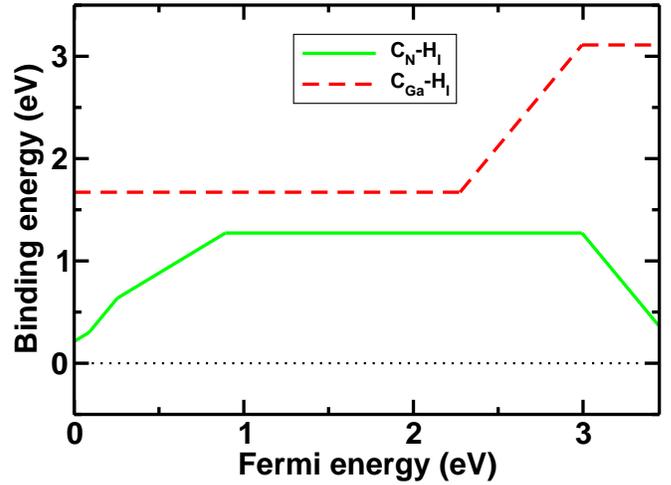}
\par\end{centering}
\caption{The calculated binding energy for $\mathrm{C_{N}-H_{I}}$ (green solid
line) and $\mathrm{C_{Ga}-H_{I}}$ (red dashed line) complexes as
a function of Fermi energy.\label{f:EbCNHi}}
\end{figure}

Formation energies of $\mathrm{C_{N}-H_{I}}$ and
$\mathrm{C_{Ga}-H_{I}}$ complex are reported with a green and red
solid lines, respectively, in FIG.~\ref{f:EfCNHi}, with those of the
$\mathrm{C_{N}}$ with a blue dashed line, $\mathrm{C_{Ga}}$ with
a cyan dashed line, $\mathrm{C_{N}-C_{Ga}}$
with a magenta dashed line, and $\mathrm{H_{I}}$ with a black solid
line as references. Both complexes have only two charge states with
energies within the band gap. For the $\mathrm{C_{N}-H_{I}}$
complex, in the very narrow region up to 0.09\,eV above the VBM, the
1+ charge state is the most stable. Above 0.09\,eV, the 0 (neutral)
charge state is the most stable up to the CBM. This clearly suggests
that the $\mathrm{C_{N}}$ (acceptor on the whole) is compensated by
the H atom (donor on the whole). In addition, this
$\mathrm{C_{N}-H_{I}}$ complex has a lower formation energy than
$\mathrm{C_{N}-C_{Ga}}$ complex, which is the most stable complex
formed by two carbons in the $n$-type region~\cite{MatsubaraC1}, by
more than 2\,eV (3\,eV) in N-rich (Ga-rich) conditions. Both in the
1+ and the 0 (neutral) charged cases the most stable positions of H
atom is the $\mathrm{AB_{\perp}}$ position. The distance between C
and H is 1.10\,\AA\ in both charge states. The structure is shown in
FIG.~\ref{f:CNHi} (a). The calculated binding energy of this complex
is plotted in FIG.~\ref{f:EbCNHi} with green solid line. The binding
energy is always positive with the value up to 1.27\,eV. Close to
both band edges the value drops less than 0.5\,eV and the complex
becomes relatively unstable.

For the $\mathrm{C_{Ga}-H_{I}}$ complex, the 2+ charge state is the
most stable when the Fermi energy is between 0 and 2.27\,eV above
the VBM\@. The neutral charge state becomes the most stable above
2.27\,eV. In the $n$-type region (upper half of the band gap) the
$\mathrm{C_{Ga}-H_{I}}$ complex has higher formation energy than the
$\mathrm{C_{N}-H_{I}}$ complex both in N-rich and Ga-rich
conditions. Comparing to the $\mathrm{C_{N}-C_{Ga}}$ complex, the
$\mathrm{C_{Ga}-H_{I}}$ complex has about 1\,eV lower formation
energy in N-rich condition, while in Ga-rich condition the formation
energies of these two complexes are almost the same. Unlike in the
$\mathrm{C_{N}-H_{I}}$ complex, the H atom takes the bond center
positions: BC$_{\perp}$ in case of the 2+ charge state, and
BC$_{\parallel}$ in case of the neutral charge state. Their
structural forms are shown in FIGs.~\ref{f:CNHi} (b) and (c),
respectively. In the BC$_{\perp}$ position, the distance between H
and C is 1.85\,\AA, whereas that between H and the nearest neighbor
N is 1.01\,\AA. In the BC$_{\parallel}$ position, the distance
between H and C is 1.08\,\AA, whereas that between H and the nearest
neighbor N is 1.71\,\AA. The binding energy of this complex, that is
shown in FIG.~\ref{f:EbCNHi} with red dashed line, is always
positive and the value increases close to the CBM to more than
3\,eV.
It should be noted that in $p$-type GaN both $\mathrm{C_{Ga}}$
and $\mathrm{H_I}$ are positively charged and expected to repel each
other. This may impede the formation of this complex.

\subsection{Silicon impurity\label{ss:silicon}}

In this section we present results obtained for silicon. First,
silicon is studied as a single impurity form (both substitutional
and interstitial). Subsequently, their complexes with carbon are
considered.

\subsubsection{Single Silicon Impurity\label{sss:Si}}

\begin{figure}
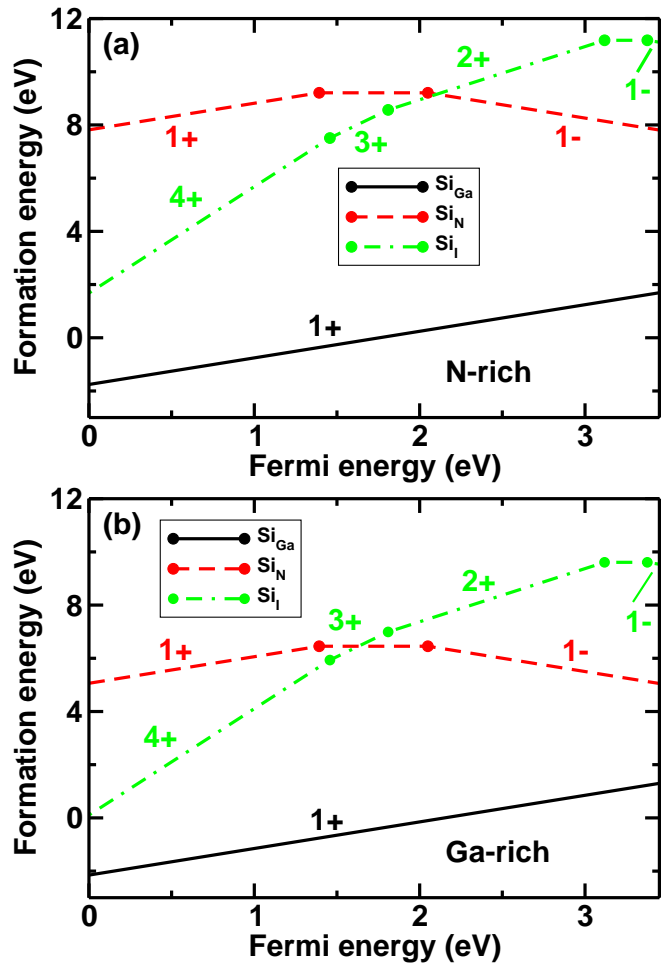

\begin{centering}
\includegraphics[width=\columnwidth]{fSii-Nrich}

\includegraphics[width=\columnwidth]{fSii-Garich}
\par\end{centering}
\caption{Formation energies as a function of Fermi energy for $\mathrm{Si_{Ga}}$
(black solid line), $\mathrm{Si_{N}}$ (red dashed line) and $\mathrm{Si_{I}}$
(green dashed-dotted line) in (a) N-rich and (b) Ga-rich conditions.\label{f:EfSi}}
\end{figure}

First we present the results obtained for single Si impurity in GaN.
As in the case of carbon shown in Part I of our
work~\cite{MatsubaraC1}, we considered Si substituting gallium
($\mathrm{Si_{Ga}}$), Si substituting nitrogen ($\mathrm{Si_{N}}$)
and interstitial Si ($\mathrm{Si_{I}}$). The calculated formation
energies, both in N-rich and Ga-rich conditions, are given in
Fig.~\ref{f:EfSi}.

For $\mathrm{Si_{Ga}}$, the 1+ charge state is the most stable within
the entire band gap (black solid line in Fig.~\ref{f:EfSi}) and
behaves as a shallow donor. This result is consistent with previous
LDA-based calculations by different groups~\cite{neugebauer96,mattila97,boguslawski97}.
In this $\mathrm{Si_{Ga}^{1+}}$ configuration, the neighboring N
atoms around Si relax slightly inward. The average Si--N bond length
is 1.78\,\AA\ ($\sim$9~\% shorter than Ga--N bond length).

$\mathrm{Si_{N}}$ has significantly higher formation energy than
$\mathrm{Si_{Ga}}$ and shows amphoteric behavior with two transition
levels around the middle of the band gap. The (+/0) donor level
appears at 1.39\,eV above the VBM and the (0/$-$) acceptor level
appears at 2.05\,eV above the VBM. Because of the size difference
between Si and N, a larger lattice deformation is observed in
contrast to the $\mathrm{Si_{Ga}}$ case. The Si--Ga bonds are
elongated in all charge states and their average lengths for the 1+,
0, $1-$ charge states are 2.29, 2.24 and 2.19\,\AA, respectively.

$\mathrm{Si_{I}}$ mostly behaves as a donor, but shows amphoteric
behavior with the $1-$ charge state being the most stable between
3.38\,eV above the VBM and the CBM. The 0 (neutral) charge state is
the most stable between 3.12\,eV and 3.38\,eV. From the VBM to
1.46\,eV above the VBM, the 4+ charge state is the most stable,
followed by the 3+ charge states between 1.46\,eV and 1.81\,eV.
subsequently, the 2+ charge state becomes the most stable between
1.81\,eV and 3.12\,eV.

\begin{figure}
\begin{centering}
\includegraphics[width=\columnwidth]{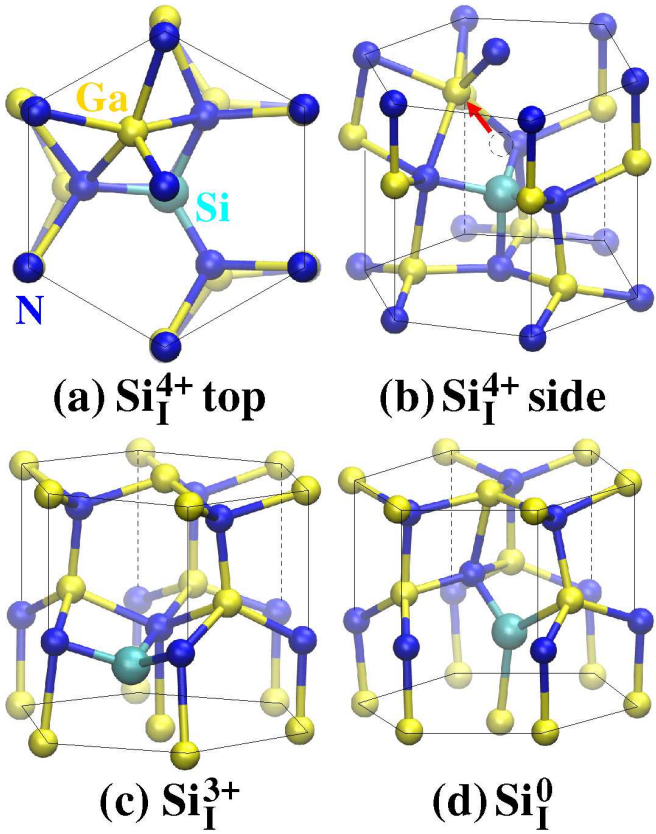}
\par\end{centering}
\caption{Ball and stick representations of the stable configurations of $\mathrm{Si_{I}}$.
Si atom is denoted by cyan sphere, while Ga and N are denoted by yellow
and blue spheres, respectively. Fully relaxed (a) tetrahedral configuration
with the 4+ charge state from the top, (b) the same from the side,
(c) octahedral configuration with the 3+ charge state and (d) type
2 split interstitial configurations.\label{f:Sii}}
\end{figure}

We considered the same starting configurations for $\mathrm{Si_{I}}$
as those of $\mathrm{C_{I}}$~\cite{MatsubaraC1}, i.e.\ octahedral,
tetrahedral, split and bond center positions. After full relaxations
we found that $\mathrm{Si_{I}}$ takes only three stable
configurations: tetrahedral (4+), octahedral (3+ and 2+) and split
type 2 (0 and $1-$) interstitial configurations. These three
configurations are shown in Fig.~\ref{f:Sii}. In the 4+ charge state
{[}Figs.~\ref{f:Sii} (a) and (b){]}, Si atom takes the tetrahedral
interstitial position and has four bonds with surrounding N atoms
(1.76\,\AA\ bond length on average). The nearest neighbor Ga atom is
pushed by the Si atom to the hexagonal channel and takes the
octahedral-like interstitial position. This movement is shown by the
red arrow in Fig~\ref{f:Sii} (b). The distance between the pushed Ga
and Si is 2.48\,\AA, whereas those between the Ga and surrounding
three N atoms are 2.07\,\AA\ on average. In the 3+ and 2+ charge
states, Si atom takes the octahedral interstitial position, where Si
atom is surrounded by three N atoms, as shown in Fig.~\ref{f:Sii}
(c). The average bond length between Si and N atoms is 1.75\,\AA\
and 1.81\,\AA\ in the 3+ and 2+ charge states. In the 0 (neutral)
and $1-$ charge states, Si atom takes the type-2 split interstitial
configuration, where tilted Si--N dimer replaces N atom and N is
positioned higher than Si [see Fig.~\ref{f:Sii} (d)]. The N atom has
two Ga--N bonds perpendicular to the $c$-axis. The Si atom also has
two bonds with Ga, but one is parallel to the $c$-axis and the other
is perpendicular to it. The average lengths of two Ga--N bonds in
the 0 and $1-$ charge states are 1.90 and 1.87\,\AA, respectively.
The Si--Ga bonds in perpendicular and parallel directions are 2.18
and 2.21\,\AA\ in the 0 charge state and 2.17 and 2.17\,\AA\ in the
$1-$ charge state, respectively.

As a single impurity form of Si, $\mathrm{Si_{Ga}}$ is the dominant
form, both in N-rich and Ga-rich conditions, with much lower
formation energies than both $\mathrm{Si_{N}}$ and
$\mathrm{Si_{I}}$. In particular, $\mathrm{Si_{I}}$ has very high
formation energy in the upper half of the band gap ($n$-type
region). Therefore, the next subsection of the manuscript will focus
on the complexes involving $\mathrm{Si_{Ga}}$ and $\mathrm{Si_{N}}$.

\subsubsection{Complexes of Silicon with Carbon\label{sss:CSi}}

\begin{figure}
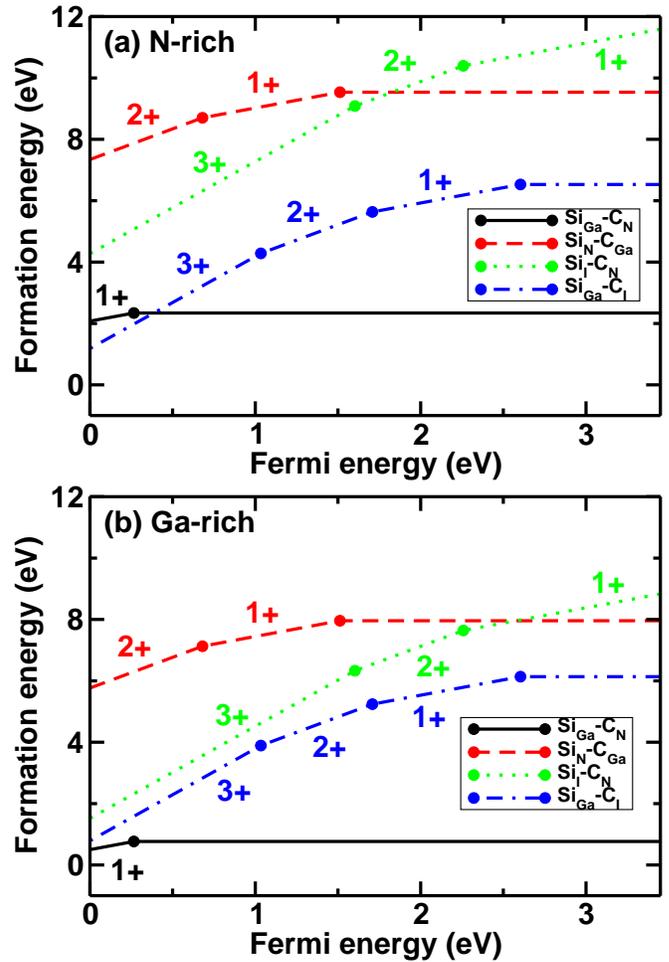

\begin{centering}
\includegraphics[width=\columnwidth]{fSiC-Nrich2}

\includegraphics[width=\columnwidth]{fSiC-Garich2}
\par\end{centering}
\caption{Formation energies as a function of Fermi energy both in (a) N-rich
and (b) Ga-rich conditions for Si--C complexes: $\mathrm{Si_{Ga}-C_{N}}$
in black solid line, $\mathrm{Si_{N}-C_{Ga}}$ in red dashed line,
$\mathrm{Si_{I}-C_{N}}$ in green dotted line and $\mathrm{Si_{Ga}-C_{I}}$
in blue dashed-dotted line.\label{f:EfSiC}}
\end{figure}

In this subsection we outline the results obtained for complexes
made of Si and C. First we deal with pairs of substitutional
impurities, i.e.\ $\mathrm{Si_{Ga}-C_{N}}$ and
$\mathrm{Si_{N}-C_{Ga}}$. Subsequently, we consider complexes made
of substitutional--interstitial pairs, i.e.\ $\mathrm{Si_{I}-C_{N}}$
and $\mathrm{Si_{Ga}-C_{I}}$. The formation energies for these Si--C
complexes are shown in Fig.~\ref{f:EfSiC} both in (a) N-rich and (b)
Ga-rich conditions.

\begin{figure}
\begin{centering}
\includegraphics[width=\columnwidth]{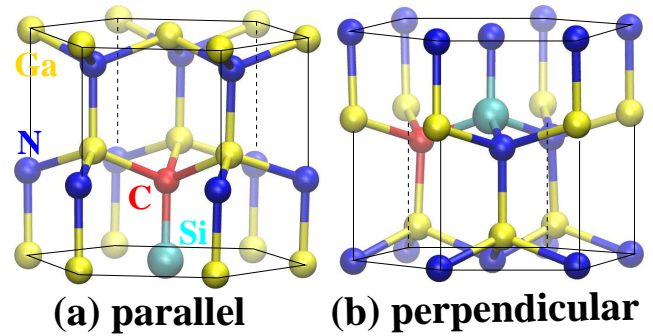}
\par\end{centering}
\caption{(a) Parallel and (b) perpendicular configurations of Si--C complex
in the case of $\mathrm{Si_{Ga}-C_{N}}$. Si atom is denoted by cyan
sphere, whereas C atom is denoted by red sphere.\label{f:sigacn}}
\end{figure}

There are two possible configurations for $\mathrm{Si_{Ga}-C_{N}}$
and $\mathrm{Si_{N}-C_{Ga}}$. In one configuration Si and C have a
bond parallel to the $c$-axis (parallel configuration) and in the
other configuration the bond is perpendicular to the $c$-axis
(perpendicular configuration). These configurations are given in
Fig.~\ref{f:sigacn}. In all the charge states except for 1+ in the
$\mathrm{Si_{Ga}-C_{N}}$ complex, the perpendicular configuration
has lower energy than the parallel configuration, but the energy
difference is very small (up to 0.11\,eV\@). The
$\mathrm{Si_{Ga}-C_{N}}$ complex has the lowest formation energy
among them. In the very small region close to the VBM, the 1+ charge
state is favorable up to 0.27\,eV above the VBM. Above that energy,
only the 0 (neutral) charge state becomes favorable throughout the
band gap. This neutral charge states is a result of the compensation
between $\mathrm{Si_{Ga}^{+}}$ and $\mathrm{C_{N}^{-}}$. The Si--C
bond lengths in the 1+ and 0 charge states are 1.85\,\AA\ and
1.81\,\AA, respectively. The $\mathrm{Si_{N}-C_{Ga}}$ complex has
three charge states within the band gap. From the VBM to 0.68\,eV
above the VBM, the 2+ charge state is favorable. Then the 1+ charge
state becomes favorable between 0.68\,eV and 1.51\,eV. Finally, the
0 (neutral) charge state becomes favorable in the rest of the band
gap. The Si--C bond lengths in the 2+, 1+ and 0 charge states are
1.84\,\AA, 1.85\,\AA\ and 1.86\,\AA, respectively.

\begin{figure}
\begin{centering}
\includegraphics[width=\columnwidth]{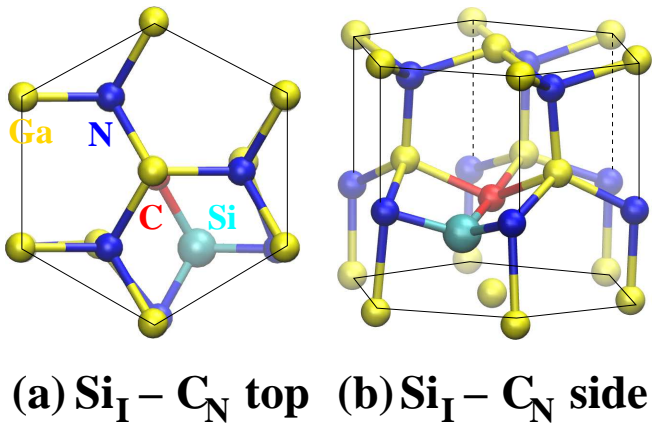}
\par\end{centering}
\caption{Relaxed structure for $\mathrm{Si_{I}-C_{N}}$ complex: (a) top view
and (b) side view. Si atom is denoted by cyan sphere, whereas C atom
is denoted by red sphere.\label{f:siicn}}
\end{figure}

$\mathrm{Si_{I}-C_{N}}$ has three stable charge states within the
band gap. The 3+ charge state is the most stable between the VBM and
1.60\,eV above the VBM. Then between 1.60\,eV and 2.25\,eV, the 2+ charge
state becomes the most stable. Finally the 1+ charge state is most
stable above 2.25\,eV. Thus, this complex behaves as a shallow donor.
In each charge state, the interstitial Si atom occupies the octahedral
interstitial position located at the middle of the hexagonal channel,
which is shown in Fig.~\ref{f:siicn}. The Si--C bond length is
1.75, 1.80 and 1.87\,\AA\ in the 3+, 2+ and 1+ charge states, respectively.
The more positive charge state the complex takes, the shorter the
bond length becomes.

\begin{figure}
\begin{centering}
\includegraphics[width=\columnwidth]{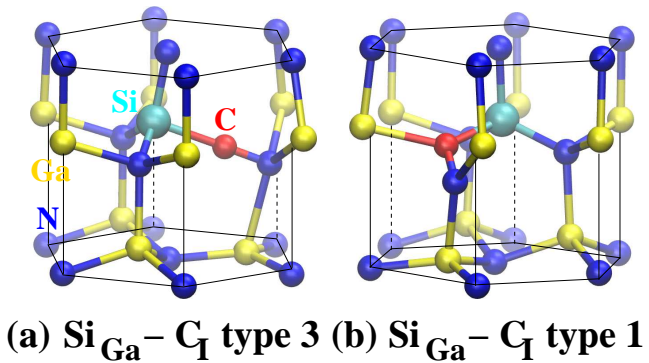}
\par\end{centering}
\caption{Relaxed structure for $\mathrm{Si_{Ga}-C_{N}}$ complex: (a) $\mathrm{C_{I}}$
forms a type 3 split interstitial with N at the site next to $\mathrm{Si_{Ga}}$
and (b) $\mathrm{C_{I}}$ forms a type 1 split interstitial with N
at the site next to $\mathrm{Si_{Ga}}$.\label{f:sigaci}}
\end{figure}

$\mathrm{Si_{Ga}-C_{I}}$ complex takes three charge states within
the band gap. When the Fermi energy is located between the VBM and
1.03\,eV above from the VBM, the complex takes the 3+ charge states
as the most stable form. Between 1.03\,eV and 1.71\,eV above the
VBM, the 2+ charge state is the most stable. Above 1.71\,eV, the
1+ charge state is the most stable up to 2.60\,eV. Finally, the neutral
charge state is the most stable above 2.60\,eV. This complex acts
as a deep donor. This complex takes two different configurations.
In both configurations, C atom is located at the site next to $\mathrm{Si_{Ga}}$
and forms a split interstitial with N atom, i.e.\ forming a C--N
dimer. In case of the 3+ charge state, the C atom forms a type 3 split
interstitial sharing the site with an N atom making a dimer, which
is defined in Ref.~\onlinecite{MatsubaraC1}. This configuration is shown in Fig.~\ref{f:sigaci}
(a). The distance between Si and C is 1.84\,\AA. In the other charge
states, the C atom forms a type 1 split interstitial, which is shown
in Fig.~\ref{f:sigaci} (b). The distance between Si and C is 1.86\,\AA,
1.79\,\AA\ and 1.78\,\AA\ for the 2+, 1+ and 0 (neutral) charge states,
respectively.

\begin{figure}
\begin{centering}
\includegraphics[width=\columnwidth]{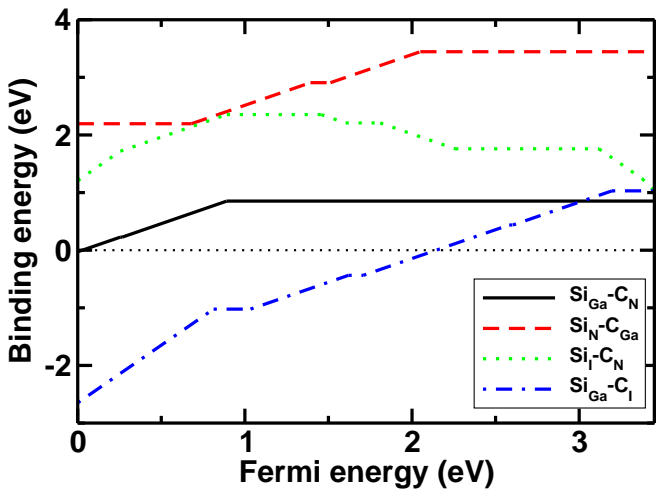}
\par\end{centering}
\caption{Binding energies as a function of Fermi energy for $\mathrm{Si_{Ga}-C_{N}}$ (black
solid line), $\mathrm{Si_{N}-C_{Ga}}$ (red dashed line), $\mathrm{Si_{I}-C_{N}}$ (green
dotted line) and $\mathrm{Si_{Ga}-C_{I}}$ (blue dashed-dotted line).\label{f:EbSiC}}
\end{figure}

The binding energies for these four different silicon-based complexes
are shown in Fig.~\ref{f:EbSiC}. For the $\mathrm{Si_{Ga}-C_{N}}$
complex (black solid line), the binding energy is negative or small
when the Fermi energy is very close to the VBM. Thus the complex is
unstable in $p$-type GaN. However, the binding energy becomes 0.85\,eV
when the Fermi energy is 0.89\,eV or above. Thus the complex is expected
to be stable in $n$-type GaN. The $\mathrm{Si_N-C_{Ga}}$ complex shows
the highest value of the binding energy (dashed red line) among the
four complexes with at least 2\,eV and up to 3.45\,eV near the CBM.
The $\mathrm{Si_{I}-C_{N}}$ complex also shows the positive binding
energy with at least 1\,eV value. Finally, the $\mathrm{Si_{Ga}-C_{I}}$
complex shows different behavior than other three complexes. When
the Fermi energy is below 2.14\,eV, the binding energy is negative.
This implies that the 3+ and the 2+ charge states are unstable as
a complex. Then the binding energy increases and reaches 1.03\,eV at
the CBM.

\subsection{Oxygen Impurity\label{ss:oxygen}}

\subsubsection{Single Oxygen Impurity\label{sss:O}}

\begin{figure}
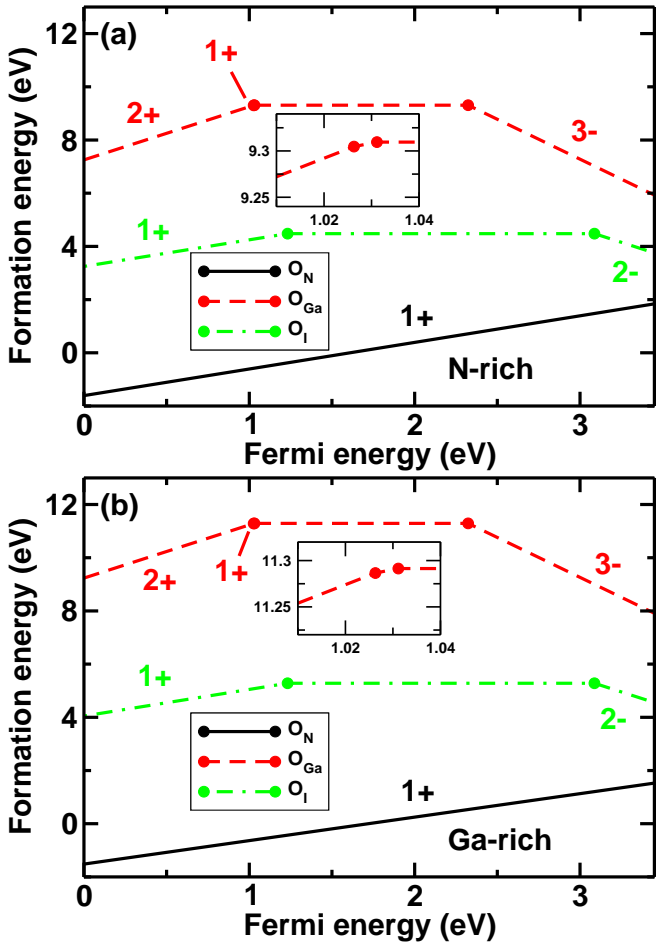

\begin{centering}
\includegraphics[width=\columnwidth]{fO-Nrich-muo}

\includegraphics[width=\columnwidth]{fO-Garich-muo}
\par\end{centering}
\caption{Formation energies as a function of Fermi energy for $\mathrm{O_{N}}$
(black solid line), $\mathrm{O_{Ga}}$ (red dashed line) and $\mathrm{O_{I}}$
(green dashed-dotted line) in (a) N-rich and (b) Ga-rich conditions.
The insets are to show the appearance of the 1+ charge state in the very
narrow region in the case of $\mathrm{O_{Ga}}$.\label{f:EfO}}
\end{figure}

As in the case of carbon and silicon, we considered three different
types of single oxygen impurity, i.e.\ O substituting Ga ($\mathrm{O_{Ga}}$),
O substituting N ($\mathrm{O_{N}}$) and interstitial O ($\mathrm{O_{\mathrm{I}}}$).
We present the formation energies of each case in Fig.~\ref{f:EfO}.

$\mathrm{O_{N}}$ acts as a shallow donor with the 1+ charge state
being most stable configuration with energy level within the band
gap. Its formation energy is very close to that of
$\mathrm{Si_{Ga}}$ and about 3 $\sim$ 5\,eV lower than that of
$\mathrm{C_{Ga}}$. The O--Ga bond lengths (2.03\,\AA\ on average)
are slightly longer than the Ga--N bonds of bulk GaN.

\begin{figure}
\begin{centering}
\includegraphics[width=\columnwidth]{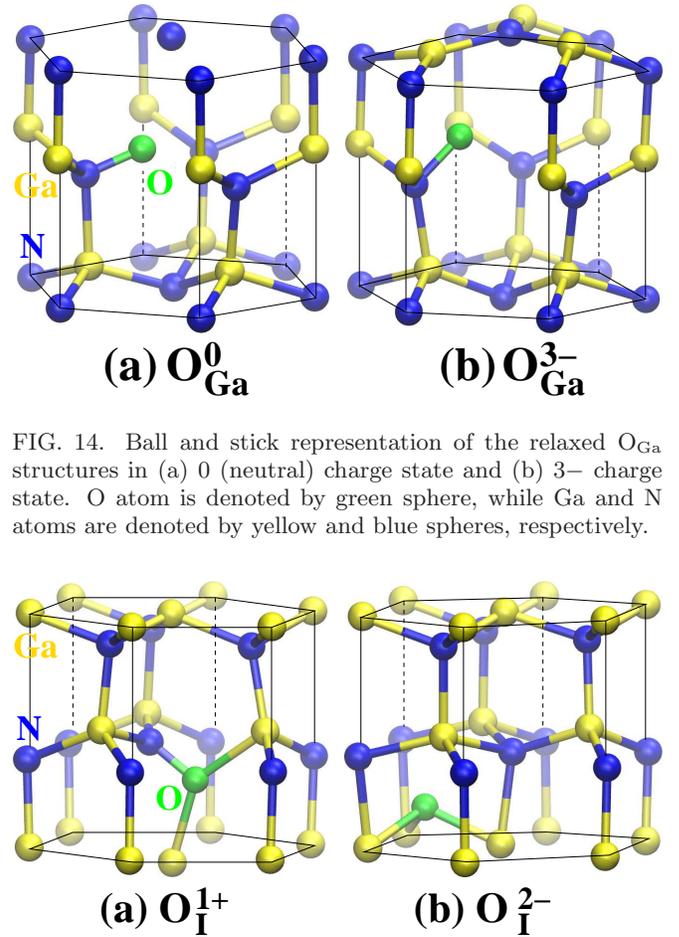}
\par\end{centering}
\caption{Ball and stick representation of the relaxed $\mathrm{O_{Ga}}$ structures
in (a) 0 (neutral) charge state and (b) $3-$ charge state. O atom
is denoted by green sphere, while Ga and N atoms are denoted by yellow
and blue spheres, respectively.\label{f:Oga}}
\end{figure}

In the case of $\mathrm{O_{Ga}}$, the 2+ charge state is the most
stable up to 1.026\,eV above the VBM. In the very small range (from
1.026\,eV to 1.031\,eV from the VBM), the 1+ charge state is the
most stable, before the 0 (neutral) charge state becomes the most
stable between 1.031\,eV and 2.32\,eV. Finally above 2.32\,eV, the
$3-$ charge state becomes the most stable state up to the CBM. In
all above mentioned charge states, O atom is not located at the Ga
atom site, but slightly displaced, making a bond with one of
surrounding N atoms. This structure is consistent with the
previously reported $\mathrm{O_{Ga}}$ structure optimized within GGA
calculations~\cite{wright05}. The structures are similar in the 2+,
1+ and 0 charge states, where the O atom is displaced by about
0.67\,\AA\ (on average) from the Ga site and forms a bond with one
of surrounding N atom with 1.26\,\AA\ (on average) bond length. As a
representative case, the structure in the 0 charge state is given in
Fig.~\ref{f:Oga}(a). The structure in the $3-$ charge state is
different from the ones in other charge states. The O atom is more
displaced from the Ga site (1.46\,\AA). It is positioned above the
Ga-layer and forms an O--N bond with 1.46\,\AA\ length. In addition,
it pushes up an N atom located at the other side of the O--N bond
{[}see Fig.~\ref{f:Oga} (b){]}, and the pushed N atom is located at
Ga-layer.

\begin{figure}
\begin{centering}
\includegraphics[width=\columnwidth]{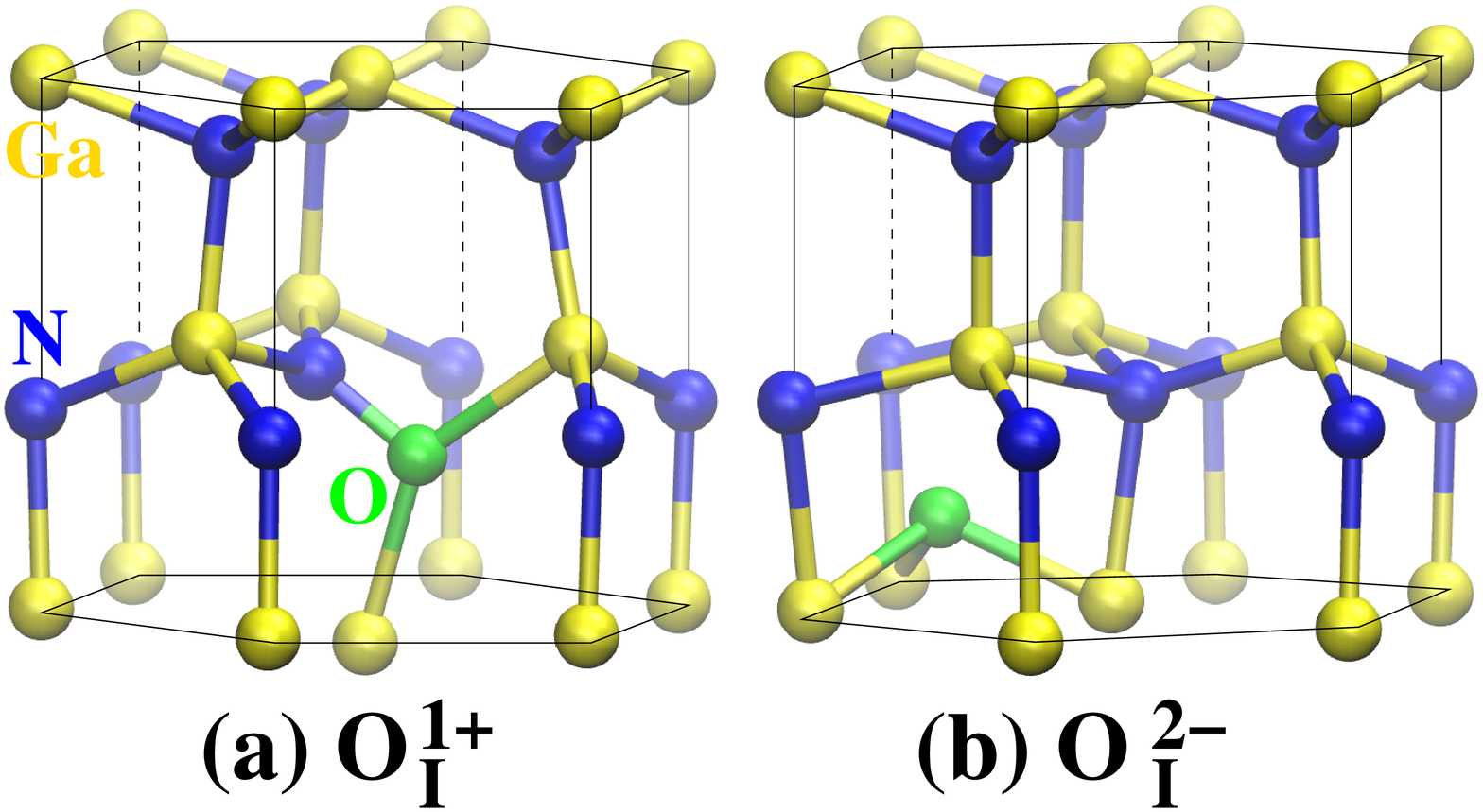}
\par\end{centering}
\caption{Ball and stick representation of the relaxed $\mathrm{O_{I}}$ structures
in the (a) 1+ charge state and (b) 2- charge state. O atom is denoted
by green sphere, while Ga and N atoms are denoted by yellow and blue
spheres, respectively.\label{f:OI}}
\end{figure}

Three charge states are observed within the band gap as stable configurations
for $\mathrm{O_{I}}$. The 1+ charge state is the most stable up to
1.23\,eV above the VBM. Then the neutral charge state becomes the
most stable up to 3.09\,eV. Finally the 2$-$ charge state is the
most stable between 3.09\,eV and the CBM. In the 1+ and neutral charge
states, O atom forms a split interstitial with a N atom. The structure
for the 1+ charge state is shown in Fig.~\ref{f:OI} (a). The distance
between O and N atom is 1.33\,\AA\ in the 1+ charge state and 1.43\,\AA\
in the neutral charge state, respectively. In the 2$-$ charge state,
O atoms is located at an octahedral interstitial position, as shown
in Fig.~\ref{f:OI} (b). The interstitial O atom has bonds with three
surrounding Ga atoms with equivalent distances of 1.88\,\AA.

\subsubsection{Complexes of Oxygen with Carbon\label{sss:CO}}

\begin{figure}
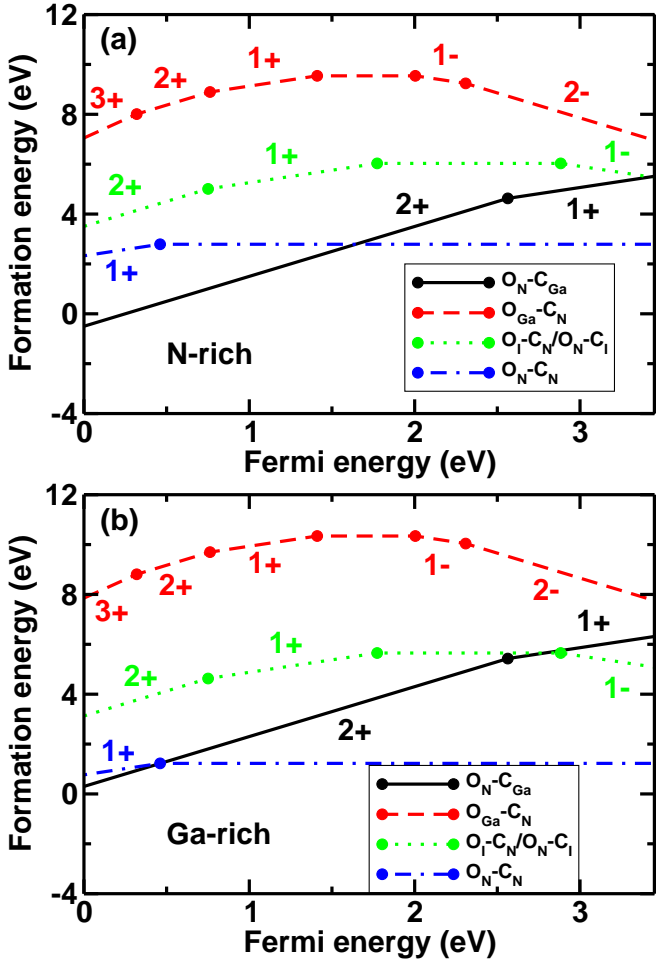

\begin{centering}
\includegraphics[width=\columnwidth]{fOC-Nrich2-muo}

\includegraphics[width=\columnwidth]{fOC-Garich2-muo}
\par\end{centering}
\caption{Formation energies as a function of Fermi energy for $\mathrm{O_{N}-C_{Ga}}$
(black solid line), $\mathrm{O_{Ga}-C_{N}}$ (red dashed line), $\mathrm{O_{I}-C_{N}}/\mathrm{O_{N}-C_{I}}$
(green dotted line) and $\mathrm{O_{N}-C_{N}}$ (blue dashed-dotted
line) in (a) N-rich and (b) Ga-rich conditions.\label{f:EfOC}}
\end{figure}

As in the case of silicon, we show the results obtained for
complexes composed of O and C. The complexes we consider here are
pairs of substitutional impurities, i.e.\ $\mathrm{O_{N}-C_{Ga}}$
and $\mathrm{O_{Ga}-C_{N}}$. We also examine complexes made of
substitutional--interstitial pairs, i.e.\ $\mathrm{O_{I}-C_{N}}$ and
$\mathrm{O_{N}-C_{I}}$, which, eventually, relax into the same
structure ($\mathrm{O_{I}-C_{N}}$) after the full geometry optimizations.
In addition to them, $\mathrm{O_{N}-C_{N}}$ complexes are also taken
into account. In this last complex, the oxygen and carbon
constituents, $\mathrm{O_{N}}$ and $\mathrm{C_{N}}$, are located as
a second nearest neighbor. This is an exceptional case, because we
have limited ourselves to consider only nearest neighbor pairs for
the constituents of complexes in all cases shown so far. Recently
this complex has been extensively studied by several groups due to
the realization that it could be the origin of the yellow
luminescence phenomena~\cite{demchenko13,lyons15,christenson15}. As
a result, we decided to include it as a potential candidate for
carbon related impurities which induce trap levels within the band
gap. The formation energies for above-mentioned O--C complexes are
shown in Fig.~\ref{f:EfOC} both in (a) N-rich
and (b) Ga-rich conditions.

In the case of $\mathrm{O_{N}-C_{Ga}}$, the 2+ and the 1+ charge
states are stable with energy levels within the band gap. The 2+
charge state is stable up to 2.56\,eV above the VBM. Above 2.56\,eV
the 1+ charge state becomes stable. Both parallel and perpendicular
configurations are considered and the perpendicular configuration
has slightly lower energies (at most 0.1\,eV) than the parallel
configuration in both charge states. This complex has relatively low
formation energy particularly close to the VBM. However, both
$\mathrm{O_{N}}$ and $\mathrm{C_{Ga}}$ are shallow donors and exist
only with positively charged states. Thus, they are expected to
repel each other under any conditions and $\mathrm{O_{N}-C_{Ga}}$
complex is unlikely to be formed despite its low formation energy.

\begin{figure}
\begin{centering}
\includegraphics[width=\columnwidth]{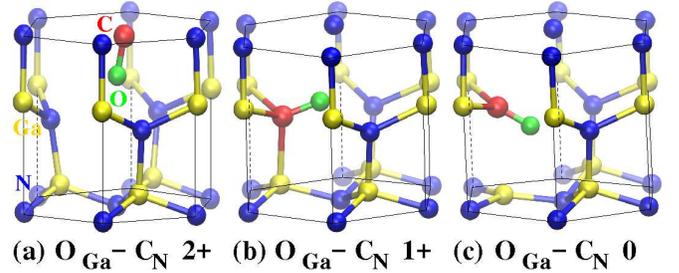}
\par\end{centering}
\caption{Ball and stick representation of the relaxed $\mathrm{O_{Ga}-C_{N}}$
structures: (a) 2+ charge state in the parallel configuration, (b)
1+ charge state in the perpendicular configuration and (c) 0 charge
state in the perpendicular configuration. O and C atoms are denoted
by green and red spheres, respectively, while Ga and N atoms are denoted
by yellow and blue spheres, respectively.\label{f:ogacn}}
\end{figure}

$\mathrm{O_{Ga}-C_{N}}$ complex takes six different charge states
from the values 3+ to 2$-$. The 3+ charge state is stable up to 0.32\,eV
above the VBM. Between 0.32\,eV and 0.76\,eV, the 2+ charge state becomes
stable. Then the stable charge state is changed to 1+ between 0.76\,eV
and 1.41\,eV. The neutral and 1$-$ charge states are favorable between
1.41\,eV and 2.00\,eV and between 2.00\,eV and 2.31\,eV, respectively.
Finally the 2$-$ state becomes stable above 2.31\,eV.
%
%
The perpendicular configuration has lower energy than the parallel
configuration in all charge states except for the 2+ case (the
difference is less than 30\,meV in this charge state, though). The
parallel configuration in the 2+ charge state is shown in
Fig.~\ref{f:ogacn}(a). $\mathrm{O_{Ga}}$ is slightly displaced from
the Ga site and is located closer to the C atom, which is located at
the N site. The O--C distance is 1.18\,\AA. In the perpendicular
configuration, O and C atoms take similar positions in the 3+, 1+,
1$-$ and 2$-$ charge states. This configuration with the 1+ charge
state is shown in Fig.~\ref{f:ogacn}(b) as a representative case. In
this case, C atom stays at the N site, while the O atom is displaced
from the Ga site toward C atom. The O--C distances are 1.13, 1.24,
1.20 and 1.26\,\AA\ in the 3+, 1+, 1$-$ and 2$-$ charge states,
respectively. In the 0 charge state, both O and C atoms are
displaced, which is shown in Fig.~\ref{f:ogacn} (c). C atom is
displaced slightly above the N site and O atom is displaced from the
Ga site toward C site and is located below the Ga site. As a result,
C atom is located higher position than O atom. The O--C distance is
1.18\,\AA.

\begin{figure}
\begin{centering}
\includegraphics[width=0.5\columnwidth]{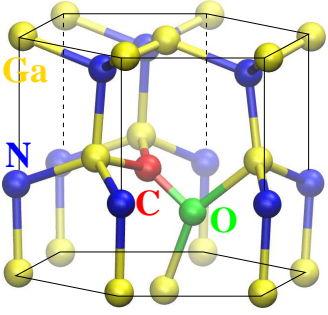}
\par\end{centering}
\caption{Ball and stick representation of the relaxed $\mathrm{O_{I}-C_{N}/O_{N}-C_{I}}$
structure in the 0 charge state. O and C atoms are denoted by green
and red spheres, respectively, while Ga and N atoms are denoted by
yellow and blue spheres, respectively.\label{f:oicn}}
\end{figure}

As already mentioned above, the $\mathrm{O_{I}-C_{N}}$ and
$\mathrm{O_{N}-C_{I}}$ complexes relax into the same structure. The resulting complex
takes the 2+, 1+, 0 and 1$-$ charge states.
The 2+ and 1$-$ charge states are favorable between the VBM
and 0.73\,eV and between 2.88\,eV and the CBM, respectively. The 1+
charge state is stable between 0.73\,eV and 1.78\,eV whereas the
neutral charge state is stable between 1.78\,eV and 2.88\,eV. In all
charge states, the C and O atoms form a dimer. The C atom always
takes higher position than the O atom. The structure is shown in
Fig.~\ref{f:oicn}. The O--C distances are 1.20, 1.28, 1.37 and
1.42\,\AA\ for 2+, 1+, 0 and 1$-$ charge states, respectively.

\begin{figure}
\begin{centering}
\includegraphics[width=\columnwidth]{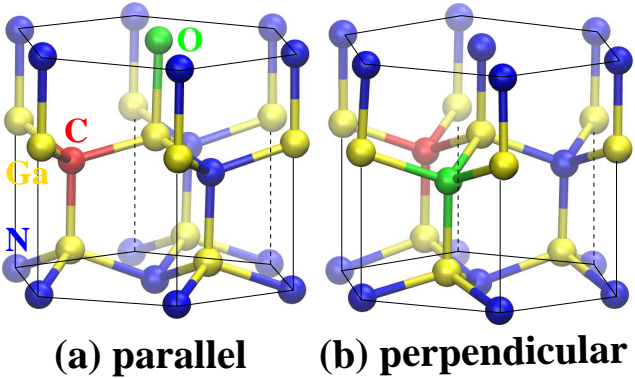}
\par\end{centering}
\caption{Ball and stick representation of the relaxed $\mathrm{O_{N}-C_{N}}$
structures with the 0 charge states in (a) parallel and (b) perpendicular
configurations. O and C atoms are denoted by green and red spheres,
respectively, while Ga and N atoms are denoted by yellow and blue
spheres, respectively. Note that O and C atoms are second nearest
neighbors to each other in this complex.\label{f:oncn}}
\end{figure}

Finally, we consider $\mathrm{O_{N}-C_{N}}$ complex. In this complex,
O and C are the second nearest neighbors to each other and two different
configurations are investigated. One is the parallel configuration,
where the O and C atoms are in the different N-plane, and the other
is the perpendicular configuration, where the O and C atoms are in
the same N-plane. These two configurations are shown in Fig.~\ref{f:oncn}
(a) and (b), respectively. The two configurations have almost the
same energy after the geometry optimizations, but the perpendicular
configurations always have slightly lower energy (less than 10\,meV difference)
than the parallel configurations in all charge states. As shown in
Fig.~\ref{f:EfOC}, this complex takes 1+ and 0 charge states with
the (+/0) transition level at 0.46\,eV. The distance between O and
C are 3.20 and 3.22\,\AA\ for the 1+ and 0 charge states, respectively.

\begin{figure}
\begin{centering}
\includegraphics[width=\columnwidth]{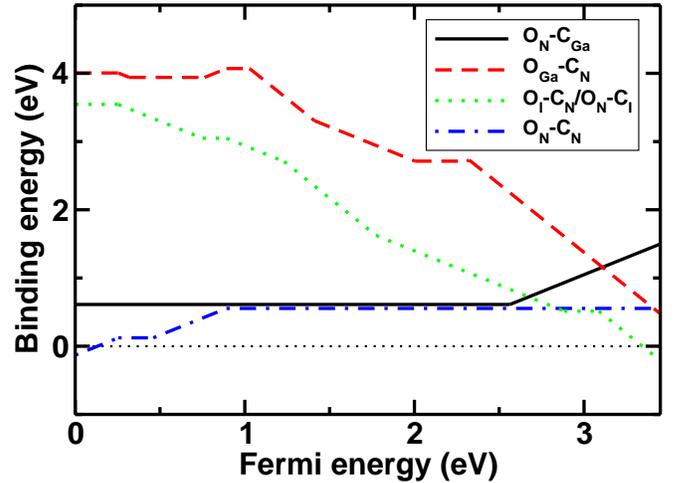}
\par\end{centering}
\caption{Binding energies as a function of Fermi energy for $\mathrm{O_{N}-C_{Ga}}$(black
solid line), $\mathrm{O_{Ga}-C_{N}}$ (red dashed line), $\mathrm{O_{I}-C_{N}/O_{N}-C_{I}}$
(green dotted line) and $\mathrm{O_{N}-C_{N}}$ (blue dashed-dotted
line).\label{f:eboc}}
\end{figure}

The binding energies for oxygen-carbon complexes studied in this
section are given in Fig.~\ref{f:eboc}. The $\mathrm{O_{N}-C_{Ga}}$
and $\mathrm{O_{N}-C_{N}}$ complexes have low binding energies.
These low binding energies values for these complexes are expected.
In the case of $\mathrm{O_{N}-C_{Ga}}$, as mentioned above, both
$\mathrm{O_{N}}$ and $\mathrm{C_{Ga}}$ act as shallow donors and
they are expected to repel each other. As for the
$\mathrm{O_{N}-C_{N}}$ complex, $\mathrm{O_{N}}$ and
$\mathrm{C_{N}}$ are second nearest neighbors, thus the binding is
expected to be weaker than the other complexes whose constituents
are first nearest neighbors. The $\mathrm{O_{Ga}-C_{N}}$ complex
shows high binding energy with 4\,eV near the VBM. Then it starts to
decrease when the Fermi energy approaches the CBM, reaching a value
of 0.48\,eV, which suggests week binding. In the case of
$\mathrm{O_{I}-C_{N}/O_{N}-C_{I}}$ complex, the binding energies are
calculated with $\mathrm{O_{I}}$ and $\mathrm{C_{N}}$ as
constituents as both $\mathrm{O_{I}-C_{N}}$ and $\mathrm{O_{N}-C_{I}}$ relax
into the same structure with the pair of $\mathrm{O_I}$ and $\mathrm{C_N}$ rather than
$\mathrm{O_N}$ and $\mathrm{C_I}$ (see Fig.~\ref{f:oicn}).
The binding energy is as high as 3.5\,eV when the Fermi energy is located
near the VBM. Then it gradually decreases and reaches zero at $E_F=3.34$\,eV.
Therefore this complex is unstable when the Fermi energy is located at 3.34\,eV
or above.

\section{Discussion\label{s:discussion}}

\subsection{Comparison with Experimental Results\label{ss:comparison}}

In this section we compare our calculated results of the carbon
related trap levels with the experimental counterparts found in
literature. Experimentally obtained carbon related trap level
energies are summarized in Table~\ref{t:exptrap}. $E_{\mathrm{TH}}$
and $E_{\mathrm{OPT}}$ denote thermal and optical activation
energies, respectively. The former is obtained by thermal techniques
such as DLTS, whereas the latter is obtained by optical techniques
such as DLOS. In our previous paper~\cite{MatsubaraC1}, we assigned
the origins of these experimentally observed trap levels to the
calculated results obtained by our HSE calculations for
carbon--carbon/carbon--vacancy complexes. These assignments are also
presented in the right column of Table~\ref{t:exptrap}.

\begin{table}
\caption{Experimentally obtained carbon related trap levels. Activation energies
obtained by thermal technique such as DLTS and by optical technique
such as DLOS are denoted by $E_{\mathrm{TH}}$ and $E_{\mathrm{OPT}}$,
respectively (in eV). Our assignments of the origins of these trap
levels based on HSE calculations for carbon--carbon/carbon--vacancy
complexes are also given.}
\label{t:exptrap}
\begin{ruledtabular}
\begin{tabular}{cc}
\multicolumn{2}{c}{Armstrong \emph{et al.}~\footnotemark[1]} \\
\fbox{$E_{c}-E_{\mathrm{OPT}}$} & \fbox{assignment} \\
$E_{c}-1.35$ & $\mathrm{C_{I}-C_{Ga}}$ \\
$E_c-3.0$ & $\mathrm{C_N}$ \\
$E_c-3.28$ & $\mathrm{C_I}$, $\mathrm{C_N-V_{Ga}}$ \\
$E_c-1.94/2.05$ & $\mathrm{C_I-C_{Ga}}$, $\mathrm{C_N-V_{Ga}}$ \\
\fbox{$E_{c}-E_{\mathrm{TH}}/E_{v}+E_{\mathrm{TH}}$} & \fbox{assignment} \\
$E_c-0.11$ & $\mathrm{C_I}$, $\mathrm{C_{Ga}-V_N}$ \\
$E_v+0.9$ & $\mathrm{C_N}$ \\
\multicolumn{2}{c}{Shah \emph{et al.}~\footnotemark[2]} \\
\fbox{$E_c-E_{\mathrm{TH}}/E_v+E_{\mathrm{TH}}$} & \fbox{assignment} \\
$E_c-0.11$ & $\mathrm{C_I}$, $\mathrm{C_{Ga}-V_N}$ \\
$E_v+0.9$ & $\mathrm{C_N}$ \\
\multicolumn{2}{c}{Polyakov \emph{et al.}~\footnotemark[3]} \\
\fbox{$E_c-E_{\mathrm{OPT}}$} & \fbox{assignment} \\
$E_c-1.3/1.4$ & $\mathrm{C_I-C_{Ga}}$ \\
$E_c-2.7/2.8$ & $\mathrm{C_I}$ \\
$E_c-3$ & $\mathrm{C_N}$ \\
\multicolumn{2}{c}{Honda \emph{et al.}~\footnotemark[4]} \\
\fbox{$E_c-E_{\mathrm{TH}}$/$E_v+E_{\mathrm{TH}}$} & \fbox{assignment} \\
$E_c-0.40$ & $\mathrm{C_I}$, $\mathrm{C_{Ga}-V_N}$ \\
$E_v+0.86$ & $\mathrm{C_N}$ \\
\end{tabular}
\end{ruledtabular}
\footnotetext[1]{Ref.~\onlinecite{armstrong05}.}
\footnotetext[2]{Ref.~\onlinecite{shah12}.}
\footnotetext[3]{Ref.~\onlinecite{polyakov13}.}
\footnotetext[4]{Ref.~\onlinecite{honda12}.}
\end{table}

Our calculated results of the trap level positions for C--H, C--Si
and C--O complexes are summarized in
Tables~\ref{t:TLH},~\ref{t:TLSi} and~\ref{t:TLO}, respectively. The
thermodynamic transition levels $\epsilon\left(q/q^{\prime}\right)$
correspond to the energy values for which the favorable charge
states are changed from $q$ to $q^{\prime}$. They are obtained using
the formation energies calculated in the previous section employing
Eq.~(\ref{eq:eqq}). Using $\epsilon\left(q/q^{\prime}\right)$, the
thermal activation energy ($E_{\mathrm{TH}}$) is expressed as
$E_{\mathrm{TH}}=E_{g}-\epsilon\left(q/q^{\prime}\right)$~\cite{MatsubaraC1}.
Since our formation energy calculations are based on the
thermodynamic equilibrium, the calculated values of
$\epsilon\left(q/q^{\prime}\right)$ and $E_{\mathrm{TH}}$ can be
directly compared with the experimental values obtained by thermal
techniques.

\begin{figure}
\begin{centering}
\includegraphics[width=\columnwidth]{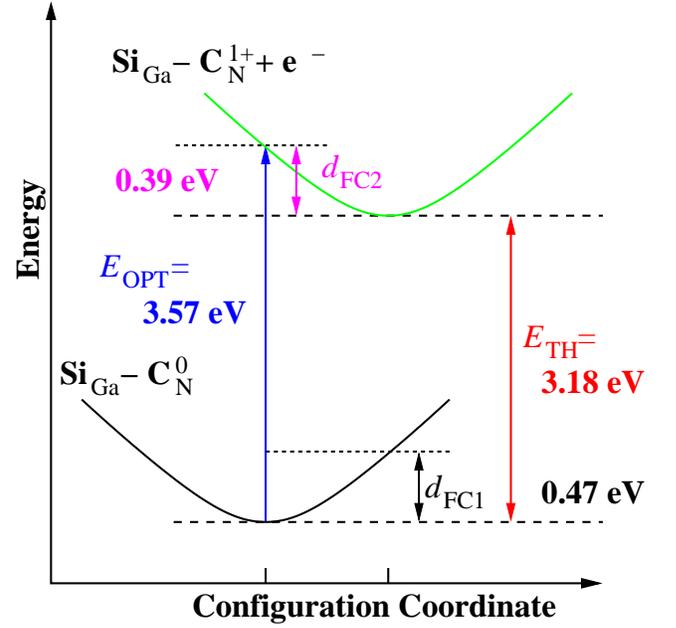}
\par\end{centering}
\caption{Schematic configuration coordinate diagram describing optical and thermal
processes by electron capture between the 1+ and 0 charge states in $\mathrm{Si_{Ga}-C_N}$
complex.\label{f:CC2}}
\end{figure}

On the other hand, in order to compare the calculated results with
the experimental trap level energies obtained by optical techniques,
we need to compute the optical activation energy
($E_{\mathrm{OPT}}$). This is obtained from $E_{\mathrm{TH}}$ with
$E_{\mathrm{OPT}} = E_{\mathrm{TH}} + d_{\mathrm{FC}}$, where
$d_{\mathrm{FC}}$ is the lattice relaxation energy (so-called
Franck-Condon shift).
In order to show the relation between $E_{\mathrm{TH}}$ and $E_{\mathrm{OPT}}$ with
the Franck-Condon shift ($d_{\mathrm{FC2}}$), schematic configuration coordinate diagram for
$\mathrm{Si_{Ga}-C_N}$ is given in Fig.~\ref{f:CC2} as a representative case.
In this complex, the $(1+/0)$ transition level was obtained as 0.27\,eV (see Fig.~\ref{f:EfSiC}).
Thus the thermal activation energy is calculated as 3.18\,eV. In order to obtain the optical
activation energy, the corresponding Franck-Condon shift ($d_{\mathrm{FC2}}$) is computed
and obtained as 0.39\,eV. This gives 3.57\,eV optical activation energy for the transition
between 1+ and 0 charge states.

\begin{table}
\caption{Thermodynamic transition levels $\left[\epsilon\left(q/q^{\prime}\right)\right]$,
thermal activation energies $\left(E_{\mathrm{TH}}\right)$ and optical
activation energies $\left(E_{\mathrm{OPT}}\right)$ for hydrogen-carbon
complexes. The levels which do not appear within the band gap are
denoted as horizontal bar. The energy levels close to the experimental
ones are denoted in bold.}
\label{t:TLH}
\begin{ruledtabular}
\begin{tabular}{ccccc}
Form & $\left(q/q^{\prime}\right)$ & $\epsilon\left(q/q^{\prime}\right)$ (eV) & $E_{\mathrm{TH}}$ (eV) & $E\mathrm{_{OPT}}$ (eV)\\ \hline
$\mathrm{C_{N}-H_{I}}$ & $\left(+/0\right)$ & 0.09 & \fbox{\textbf{3.36}} & -\\
$\mathrm{C_{Ga}-H_{I}}$ & $\left(2+/0\right)$ & 2.27 & 1.18 & -\\
 & $\left(2+/+\right)$ & 2.31 & 1.14 & \fbox{\textbf{2.13}}\\
 & $\left(+/0\right)$ & 2.23 & 1.22 & \fbox{\textbf{2.56}}\\
\end{tabular}
\end{ruledtabular}
\end{table}

Our calculated transition level positions and activation energies
for C--H complexes are reported in Table~\ref{t:TLH}. The
$\mathrm{C_{N}-H_{I}}$ complex has only one transition level, which
is $\left(+/0\right)$ and its thermal activation energy is 3.36\,eV.
This energy is close to the experimental value of 3.20\,eV observed
by Shah \emph{et al.}~\cite{shah12}. The $\mathrm{C_{Ga}-H_{I}}$
complex also has one transition level, $\left(2+/0\right)$, at
2.27\,eV, whose thermal activation energy corresponds to 1.18\,eV.
In addition, two optical transitions are possible:
$\left(2+/+\right)$ at 2.13\,eV and $\left(+/0\right)$ at 2.56\,eV.
The former may correspond to the trap at $E_{c}-1.94/2.05$\,eV
obtained by Armstrong \emph{et al.}~\cite{armstrong05}, whereas the
latter may correspond to the trap at $E_{c}-2.7/2.8$\,eV obtained by
Polyakov \emph{et al.}~\cite{polyakov13}.

\begin{table}
\caption{Thermodynamic transition levels $\left[\epsilon\left(q/q^{\prime}\right)\right]$,
thermal activation energies $\left(E_{\mathrm{TH}}\right)$ and optical
activation energies $\left(E_{\mathrm{OPT}}\right)$ for silicon-carbon
complexes. The levels which do not appear within the band gap are
denoted as horizontal bar. The energy levels close to the experimental
ones are denoted in bold.}
\label{t:TLSi}
\begin{ruledtabular}
\begin{tabular}{ccccc}
Form & $\left(q/q^{\prime}\right)$ & $\epsilon\left(q/q^{\prime}\right)$ (eV) & $E_{\mathrm{TH}}$ (eV) & $E\mathrm{_{OPT}}$ (eV)\\ \hline
$\mathrm{Si_{Ga}-C_{N}}$ & $(+/0)$ & 0.27 & \fbox{\textbf{3.18}} & --\\
$\mathrm{Si_{N}-C_{Ga}}$ & $(2+/+)$ & 0.68 & 2.77 & 3.42 \\
 & $(+/0)$ & 1.51 & 1.94 & 2.36\\
$\mathrm{Si_{I}-C_{N}}$ & $(3+/2+)$ & 1.60 & 1.85 & \fbox{\textbf{2.65}}\\
 & $(2+/+)$ & 2.26 & 1.19 & 2.28\\
$\mathrm{Si_{Ga}-C_{I}}$ & $(3+/2+)$ & 1.03 & 2.42 & --\\
 & $(2+/+)$ & 1.71 & 1.74 & 2.55\\
 & $(+/0)$ & 2.60 & 0.85 & \fbox{\textbf{1.43}}\\
\end{tabular}
\end{ruledtabular}
\end{table}

The calculated transition levels obtained for Si--C complexes are
summarized in Table~\ref{t:TLSi}. The $\left(+/0\right)$ transition
level of $\mathrm{Si_{Ga}-C_{N}}$ complex has $E_{\mathrm{TH}} =
3.18$\,eV. This energy value is close to the $E_{c}-3.20$\,eV
obtained by Shah \emph{et al.}~\cite{shah12}
The $\left(2+/+\right)$
level from the $\mathrm{Si_{N}-C_{Ga}}$ complex has
$E_{\mathrm{TH}}=2.77$\,eV and $E_{\mathrm{OPT}}=3.42$\,eV. The
former corresponds to the level at $E_{c}-2.69$\,eV obtained by Shah
\emph{et al.}\ with DLTS~\cite{shah12}, while the latter has close
energy to the level at $E_{c}-3.28$\,eV observed by Armstrong
\emph{et al.}\ with DLOS~\cite{armstrong05}.
However, the lowest formation energy for $\mathrm{Si_N-C_{Ga}}$ is 5.77\,eV (see Fig.~\ref{f:EfSiC}), which is high compared to other Si--C complexes.
Using Eq.~(\ref{eq:C}) with typical MOCVD growth temperature of 1323\,K, the concentration of this
complex is negligible ($\sim$10$^1$\,cm$^{-3}$ at most). Thus we do not assign the levels from this complex to any experimentally observed trap levels in Tables~\ref{t:TLSi} and~\ref{t:assignments}.
The $\left(3+/2+\right)$
level from the $\mathrm{Si_{I}-C_{N}}$ complex has 2.65\,eV optical
activation energy. This value is very close to the trap level at
$E_{c}-2.7/2.8$\,eV observed by Polyakov \emph{et
al.~\cite{polyakov13}. }The $\mathrm{Si_{Ga}-C_{I}}$ complex has
negative binding energy up to 2.14\,eV. Therefore only the
$\left(+/0\right)$ level is likely to exist as one from stable form
complexes. Its optical activation energy of 1.43\,eV is very close
energy to the levels observed by Armstrong \emph{et al.} at
$E_{c}-1.35$\,eV~\cite{armstrong05} and by Polyakov \emph{et al.} at
$E_{c}-1.3/1.4$\,eV~\cite{polyakov13}.

\begin{table}
\caption{Thermodynamic transition levels $\left[\epsilon\left(q/q^{\prime}\right)\right]$,
thermal activation energies $\left(E_{\mathrm{TH}}\right)$ and optical
activation energies $\left(E_{\mathrm{OPT}}\right)$ for oxygen-carbon
complexes. The levels which do not appear within the band gap are
denoted as horizontal bar. The energy levels close to the experimental
ones are denoted in bold.}
\label{t:TLO}
\begin{ruledtabular}
\begin{tabular}{ccccc}
Form & $\left(q/q^{\prime}\right)$ & $\epsilon\left(q/q^{\prime}\right)$ (eV) & $E_{\mathrm{TH}}$ (eV) & $E\mathrm{_{OPT}}$ (eV)\\ \hline
$\mathrm{O_{N}-C_{Ga}}$ & $\left(2+/+\right)$ & 2.56 & 0.89 & 1.77\\
$\mathrm{O_{N}-C_{N}}$ & $\left(+/0\right)$ & 0.46 & 2.99 & \fbox{\textbf{3.40}}\\
$\mathrm{O_{Ga}-C_{N}}$ & $\left(3+/2+\right)$ & 0.32 & 3.13 & --\\
 & $\left(2+/+\right)$ & 0.76 & 2.69 & --\\
 & $\left(+/0\right)$ & 1.41 & 2.04 & 2.30\\
 & $\left(0/-\right)$ & 2.00 & 1.45 & 1.88\\
 & $\left(-/2-\right)$ & 2.31 & 1.14 & 1.39\\
$\mathrm{O_{I}-C_{N}}/$ & $\left(2+/+\right)$ & \fbox{\textbf{0.75}} & \fbox{\textbf{2.70}} & --\\
$\mathrm{O_{N}-C_{I}}$ & $\left(+/0\right)$ & 1.77 & 1.68 & 2.35\\
 & $\left(0/-\right)$ & 2.88 & \fbox{\textbf{0.57}} & \fbox{\textbf{1.53}}\\
\end{tabular}
\end{ruledtabular}
\end{table}

Finally we perform a comparison between our calculated transition
levels for oxygen-carbon complexes and the experimentally observed
ones. The $\mathrm{O_{N}-C_{Ga}}$ complex is unlikely to form
because both constituents are shallow donors and are expected to
repel each other under any conditions. Therefore there is no
experimentally observed trap levels which would correspond to the
$\left(2+/+\right)$ trap level of the $\mathrm{O_{N}-C_{Ga}}$
complex. The optical activation energy of $\left(+/0\right)$ of the
$\mathrm{O_{N}-C_{N}}$ complex is calculated to be 3.40\,eV. This
energy is close to the experimentally observed $E_{c}-3.28$\,eV trap
level~\cite{armstrong05}. The $\left(2+/+\right)$ level of the
$\mathrm{O_{Ga}-C_{N}}$ complex is located at 0.76\,eV with 2.69\,eV
thermal activation energy. The former corresponds to the trap levels
at $E_{v}+0.9$\,eV observed by Armstrong \emph{et
al.}~\cite{armstrong05} and at $E_{v}+0.86$\,eV observed by Honda
\emph{et al.}~\cite{honda12}, whereas the latter corresponds to the
trap level at $E_{c}-2.69$\,eV measured by Shah \emph{et
al.}~\cite{shah12}. In addition, the optical activation energy of
the $\left(0/-\right)$ level (1.88\,eV) is close to the
$E_{c}-1.94/2.05$\,eV level reported by Armstrong \emph{et
al.}~\cite{armstrong05} and that of $\left(-/2-\right)$ level
(1.39\,eV) may correspond to the $E_{c}-1.35$\,eV level measured by
Armstrong \emph{et al.}~\cite{armstrong05} and the
$E_{c}-1.3/1.4$\,eV level obtained by Polyakov \emph{et
al.}~\cite{polyakov13}.
However, as in the case of $\mathrm{Si_N-C_{Ga}}$, $\mathrm{O_{Ga}-C_N}$ has high formation energy with the value at least 6.96\,eV (see Fig.~\ref{f:EfOC}).
Again, using Eq.~(\ref{eq:C}) with the temperature of 1323\,K, the concentration of this
complex is negligible ($<$ 1\,cm$^{-3}$). Thus we do not assign the levels from this complex to any experimentally observed trap levels in Tables~\ref{t:TLO} and~\ref{t:assignments}.

The $\left(2+/+\right)$ level of
$\mathrm{O_{I}-C_{N}/O_{N}-C_{I}}$ complex has the
$\left(2+/+\right)$ trap level at 0.75\,eV with 2.70\,eV thermal
activation energy, which are very similar to the $\left(2+/+\right)$
level of $\mathrm{O_{Ga}-C_{N}}$ complex. The $\left(0/-\right)$
level of $\mathrm{O_{I}-C_{N}/O_{N}-C_{I}}$ complex has 0.57\,eV
thermal activation energy and 1.53\,eV optical activation energy.
The former is close to the the trap level at $E_{c}-0.40$\,eV
observed by Honda \emph{et al.}~\cite{honda12}, while the latter is
close to the trap level at $E_{c}-1.35$\,eV observed by Armstrong
\emph{et al.}~\cite{armstrong05} and at $E_{c}-1.3/1.4$\,eV by
Polyakov \emph{et al.}~\cite{polyakov13}.

\begin{table*}
\caption{Assignments of the experimentally observed trap levels to our theoretically
obtained trap levels considering all the defects considered. Assignments based on previous
LDA/HSE calculations are also given. For experimental values, (T) denotes the level
obtained by thermal techniques, whereas (O) denotes the level obtained by optical techniques.}
\label{t:assignments}
\begin{ruledtabular}
\begin{tabular}{cccc}
experiment & previous & our HSE~\footnotemark[7] & our HSE~\footnotemark[8] \\
& LDA~\footnotemark[5]/HSE~\footnotemark[6] & C, C--C, C--V & C--H, C--Si, C--O \\ \hline
$E_{v}+0.9~\footnotemark[1],E_{v}+0.86~\footnotemark[2]$ (T) & $\mathrm{C_{N}}$ & $\mathrm{C_{N}}$ & $-$ \\
& & & $\mathrm{O_{I}-C_{N}/O_{N}-C_{I}}$ \\
$E_{c}-0.11~\footnotemark[1], E_{c}-0.13~\footnotemark[3]$ (T) & $\mathrm{C_{Ga}}$ & $\mathrm{C_{I}, C_{Ga}-V_{N}}$ & $-$\\
$E_{c}-0.40~\footnotemark[2]$ (T) & $-$ & $\mathrm{C_{I}, C_{Ga}-V_{N}}$ & $\mathrm{O_{I}-C_{N}/O_{N}-C_{I}}$\\
$E_{c}-1.35~\footnotemark[1],E_{c}-1.3/1.4~\footnotemark[4]$ (O) & $\mathrm{C_{I}}$ & $\mathrm{C_{I}-C_{Ga}}$ & $\mathrm{Si_{Ga}-C_{I}}$ \\
& & & $\mathrm{O_{I}-C_{N}/O_{N}-C_{I}}$ \\
$E_{c}-1.94/2.05~\footnotemark[1]$ (O) & $-$ & $\mathrm{C_{I}-C_{Ga}, C_{N}-V_{Ga}}$ & $\mathrm{C_{Ga}-H_{I}}$ \\
$E_{c}-2.69~\footnotemark[3]$ (T) & $\mathrm{C_{N}/V_{Ga}}$ & $\mathrm{C_{N}}$ & $-$ \\
& & & $\mathrm{O_{I}-C_{N}/O_{N}-C_{I}}$ \\
$E_{c}-2.7/2.8~\footnotemark[4]$ (O) & $-$ & $\mathrm{C_{I}}$ & $\mathrm{C_{Ga}-H_{I}}$, $\mathrm{Si_{I}-C_{N}}$ \\
$E_{c}-3.0~\footnotemark[1],E_{c}-3~\footnotemark[4]$ (O) & $\mathrm{C_{N}}$ & $\mathrm{C_{N}}$ & $-$ \\
$E_{c}-3.20~\footnotemark[3]$ (T) & $\mathrm{C_{N}}$ & $\mathrm{C_{N}}$ & $\mathrm{C_{N}-H_{I}}$, $\mathrm{Si_{Ga}-C_{N}}$ \\
$E_{c}-3.28~\footnotemark[1]$ (O) & $\mathrm{C_{N}}$ & $\mathrm{C_{I}, C_{N}-V_{Ga}}$ & $\mathrm{O_{N}-C_{N}}$
\end{tabular}
\end{ruledtabular}
\footnotetext[1]{Ref.~\onlinecite{armstrong05}.}
\footnotetext[2]{Ref.~\onlinecite{honda12}.}
\footnotetext[3]{Ref.~\onlinecite{shah12}.}
\footnotetext[4]{Ref.~\onlinecite{polyakov13}.}
\footnotetext[5]{Ref.~\onlinecite{wright02}.}
\footnotetext[6]{Ref.~\onlinecite{lyons10}.}
\footnotetext[7]{Ref.~\onlinecite{MatsubaraC1}.}
\footnotetext[8]{This work.}
\end{table*}

Table~\ref{t:assignments} summarizes our assignments of the
calculated values to the experimentally observed trap levels. The
right most column corresponds to the C complexes with H, Si or O\@.
All these unintentional impurities contribute, with different
efficacy, to the introduction of C-related trap levels in the band
gap.

Furthermore, to provide additional information for the experimental
groups, we also calculated vibrational frequencies of the complexes
with interstitials~\cite{FDmethod}, which can be
assigned to experimentally observed trap levels in order to provide
information for the detection of these complexes. The frequencies
for the stretching modes are summarized in Table~\ref{t:Vib}.

\begin{table}
\caption{Vibrational frequencies for $\mathrm{C_N-H_I}$, $\mathrm{C_{Ga}-H_I}$,
$\mathrm{Si_I-C_N}$, $\mathrm{Si_{Ga}-C_I}$ and $\mathrm{I_I-C_N/O_N-C_i}$.
The stretching mode frequencies are given in cm$^{-3}$.\label{t:Vib}}
\begin{ruledtabular}
\begin{tabular}{lll}
Form & $q$ & vibrational frequency \\ \hline
$\mathrm{C_N-H_I}$ & $1+$ & 3023 \\
               & $0$  & 3062 \\
$\mathrm{C_{Ga}-H_I}$ & $2+$ & 3656 \\
                      & $1+$ & 3041 \\
                      & $0$ & 3109 \\
$\mathrm{Si_I-C_N}$ & $3+$ & 1011 \\
                    & $2+$ & 911 \\
$\mathrm{Si_{Ga}-C_I}$ & $1+$ & 1570 \\
                       & $0$ & 1363 \\
$\mathrm{O_I-C_N}/$ & $2+$ & 1779 \\
$\mathrm{O_N-C_I}$                                    & $1+$ & 1449 \\
                                    & $0$ & 1258 \\
                                    & $1-$ & 1176 \\
\end{tabular}
\end{ruledtabular}
\end{table}

\subsection{Defect Concentration\label{ss:concentration}}

In order to experimentally detect these C-related defects, they must
have a concentration that is above the experimental measurable
threshold. Therefore, in this subsection we calculate the
concentrations of these defects to estimate the amount
of carbon effectively incorporated in GaN.

The concentration of defect $i$, $\left[C_{i}\right]$, is determined
by Eq.~(\ref{eq:C}) using defect formation energy computed by
Eq.~(\ref{eq:Ef}) and growth temperature $T$. The formation energy
is a function of the Fermi level ($E_{\mathrm{F}}$), therefore it is
a variable here and must be determined self-consistently to satisfy
the following charge neutrality condition:
\begin{eqnarray}
\sum_{i}q_{i}\left[C_{i}\right]-\left[n\right]+\left[p\right] & = & 0,\label{eq:chgneut}
\end{eqnarray}
where $q_{i}$ is the charge state, $\left[n\right]$ and $\left[p\right]$
are the electron and hole concentrations, respectively, and they are
written as
\begin{eqnarray}
\left[n\right] & = & N_{c}\exp\left(-\frac{E_{g}-E_{\mathrm{F}}}{k_{\mathrm{B}}T}\right),
\end{eqnarray}
where
\begin{eqnarray}
N_{c} & = & \frac{2\left(2\pi m_{n}^{*}k_{\mathrm{B}}T\right)^{3/2}}{h^{3}},
\end{eqnarray}
and
\begin{eqnarray}
\left[p\right] & = & N_{v}\exp\left(-\frac{E_{\mathrm{F}}}{k_{\mathrm{B}}T}\right),
\end{eqnarray}
where
\begin{eqnarray}
N_{v} & = & \frac{2\left(2\pi m_{p}^{*}k_{\mathrm{B}}T\right)^{3/2}}{h^{3}}.
\end{eqnarray}
Here, $E_{g}$ is the band gap, $m_{n}^{*}$ ($m_{p}^{*}$) is the
electron (hole) effective mass and $h$ is the Planck's constant. We
used the effective masses equal to 0.2$m_e$~\cite{drechsler95} for
electron and 0.8$m_e$~\cite{pankove75} for hole with free electron
mass $m_e$.

\begin{figure}
\begin{centering}
\includegraphics[width=\columnwidth]{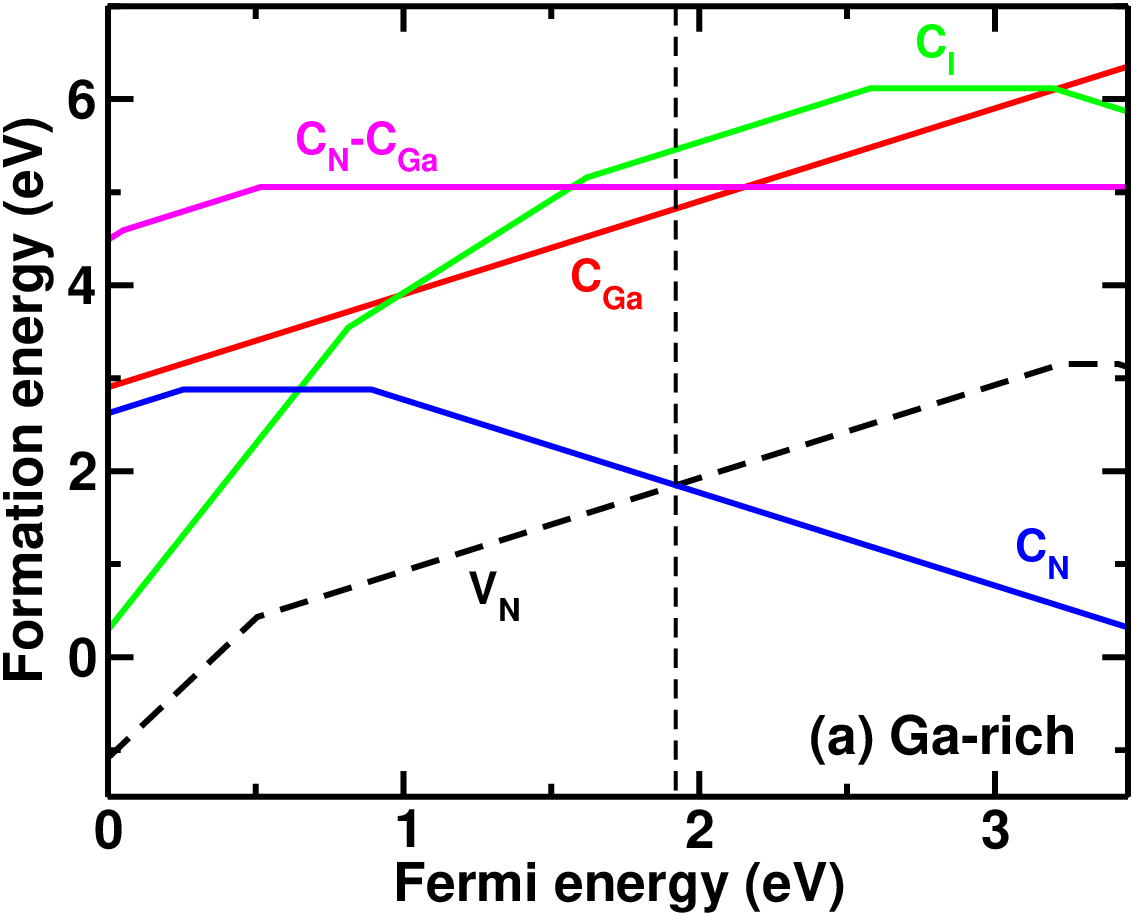}

\includegraphics[width=\columnwidth]{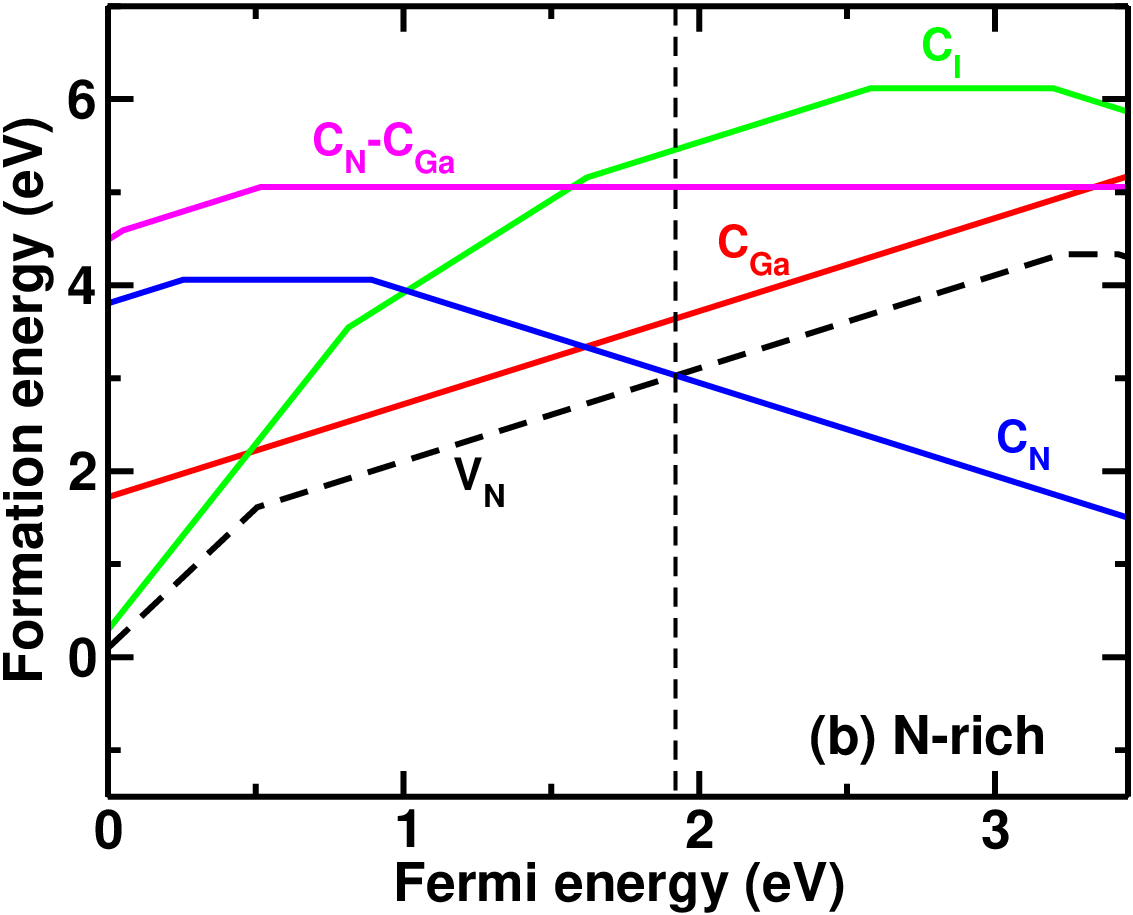}
\par\end{centering}
\caption{Formation energies as a function of Fermi energy for defects with concentrations
more than 1$\times10^{10}$\,cm$^{-3}$ in (a) Ga-rich and (b) N-rich conditions. The C-related
formation energies are plotted with solid lines, while non C-related ones are with dashed lines.
The dashed vertical bar corresponds to the equilibrium Fermi energy obtained by the charge neutrality
conditions.\label{f:EfdomCV}}
\end{figure}

\begin{figure}
\begin{centering}
\includegraphics[width=\columnwidth]{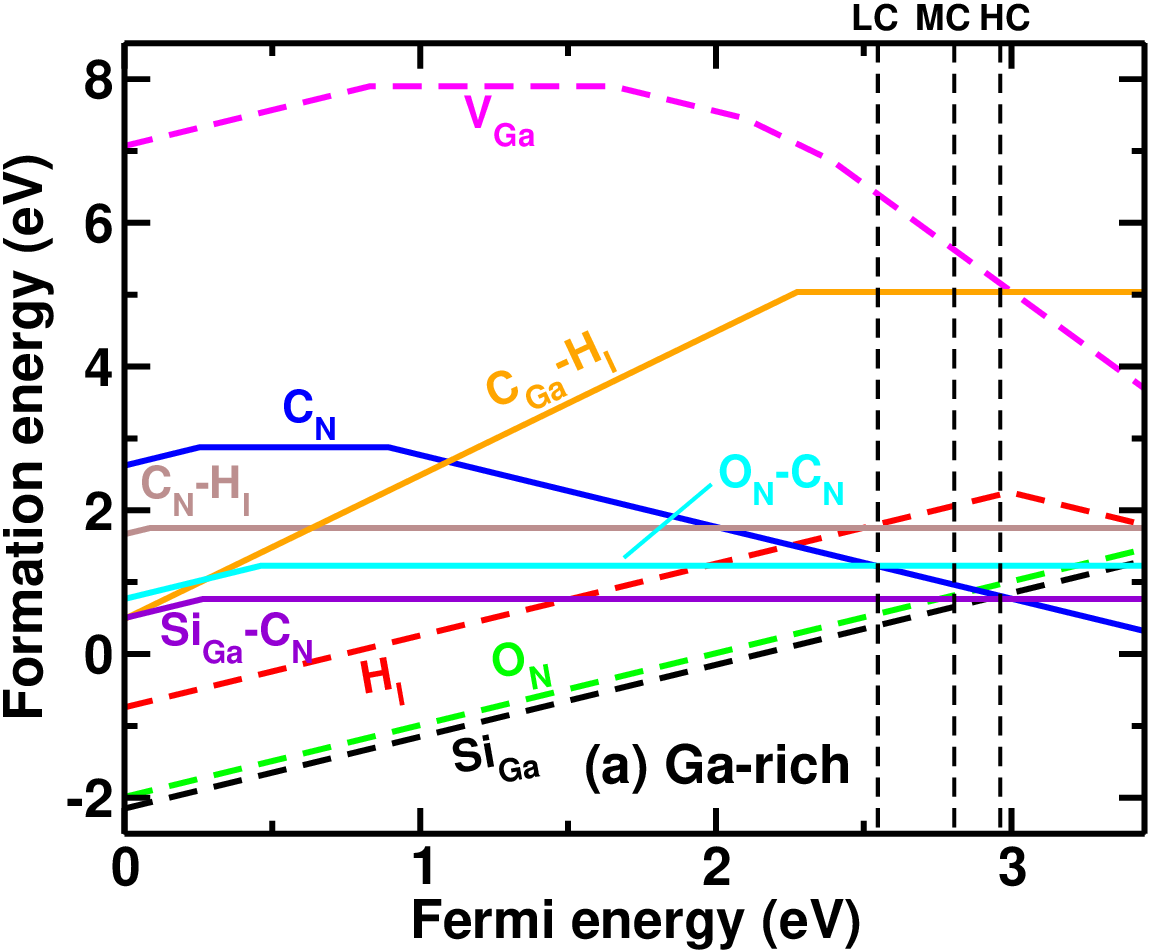}

\includegraphics[width=\columnwidth]{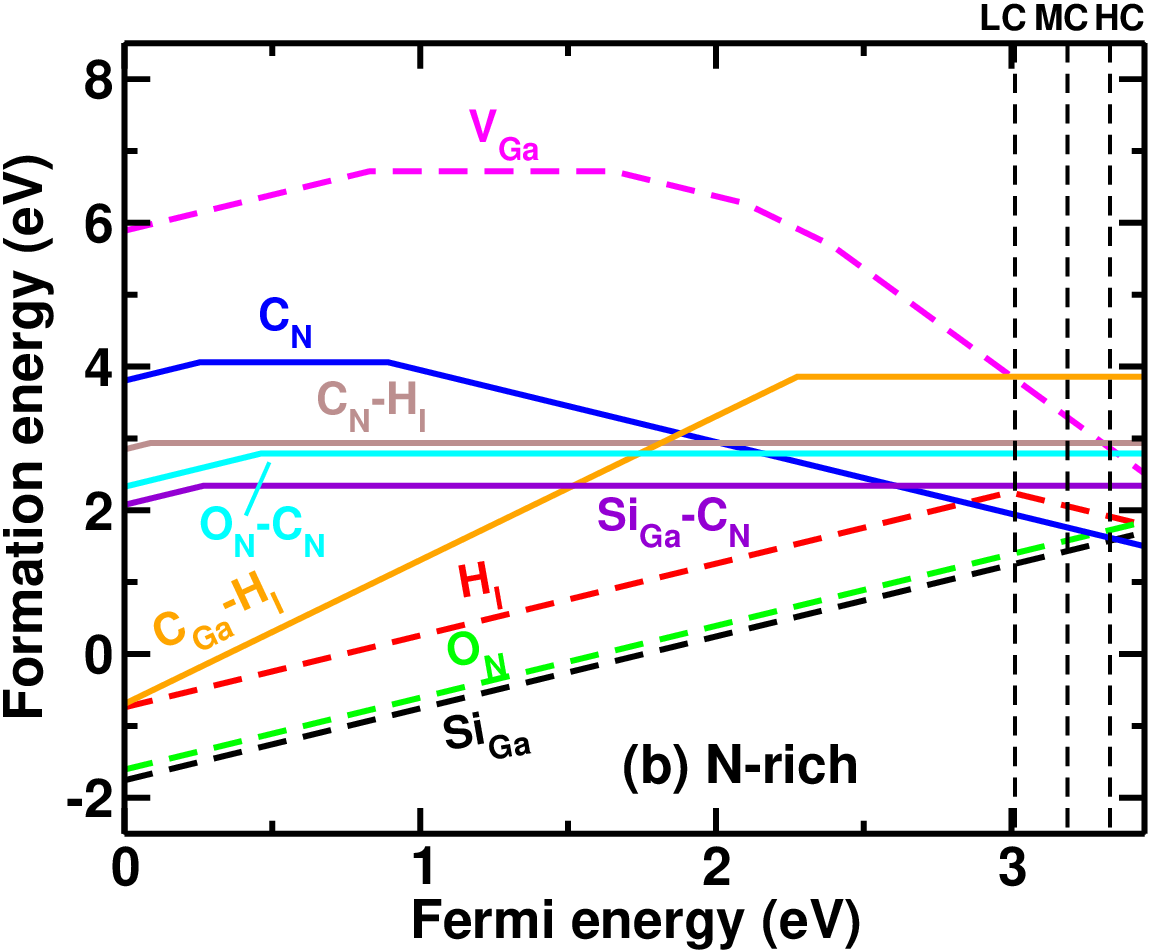}
\par\end{centering}
\caption{Formation energies as a function of Fermi energy for defects with concentrations
more than 1$\times10^{10}$\,cm$^{-3}$ in (a) Ga-rich and (b) N-rich conditions. The C-related
formation energies are plotted with solid lines, while non C-related ones are with dashed lines.
The dashed vertical bars correspond to the equilibrium Fermi energy obtained by the charge neutrality
conditions.\label{f:EfdomHSO}}
\end{figure}

In the following the band gap value ($E_g$=3.45\,eV) and the growth
temperature ($T$=1323\,K as typical MOCVD growth temperature) are
fixed to obtain equilibrium Fermi energy in solving
Eq.~(\ref{eq:chgneut}).
In order to consider different carbon concentration scenarios, we
introduce a scaling factor $\alpha$ for the carbon concentration
following the method by Wright~\cite{wright05}. We consider three
different situations, i.e. low carbon (LC,
$1\times10^{15}$\,cm$^{-3}$), mid carbon (MC,
$1\times10^{17}$\,cm$^{-3}$) and high carbon (HC,
$1\times10^{19}$\,cm$^{-3}$) concentrations. In the LC situation,
the total C concentration is expressed by
\begin{eqnarray}
1\times10^{15}\,\mathrm{cm}^{-3} &=& \alpha\sum_{j}\left[C_{j}\right]
= \alpha\left[C_{\mathrm{C_N}^{1-}}\right]
+ \alpha\left[C_{\mathrm{C_N}^{0}}\right] + \cdots, \label{eq:cconc} \nonumber\\
\end{eqnarray}
where $\left[C_{j}\right]$ denotes the concentrations for all kinds
of carbon related defects with different charge states. Under the
constraint described by Eq.~(\ref{eq:cconc}), $\alpha$ and $E_F$ are
determined by solving Eq.~(\ref{eq:chgneut}) iteratively.
We considered two different situations. In the first case (a), one
assumes that C, vacancy (V) and their complexes are the only
electrically active defects in the system. In the second case (b),
all defects considered (H, Si, O and their complexes with C as well
as C and V) are electrically active in the system.
The results are summarized in Tables~\ref{t:concCV}
and~\ref{t:concHSO}, where only defects with more than
1$\times10^{10}$\,cm$^{-3}$ concentration are shown. The formation
energies for these defects with substantial concentrations are
plotted in Figs.~\ref{f:EfdomCV} and~\ref{f:EfdomHSO} in the
respective cases.

\begin{table*}
\caption{Concentrations of C and vacancies in the system. Only the ones with more than
10$^{10}$\,cm$^{-3}$ are shown. Three different C concentrations are denoted as
LC (low C concentration), MC (medium C concentration) and HC (high C concentration).
$E_F^{\mathrm{eq}}$ denotes the equilibrium Fermi energy in eV. Highest C concentrations
in each condition are denoted in bold. All the concentrations
are in units of cm$^{-3}$. \label{t:concCV}}
\begin{ruledtabular}
\begin{tabular}{ccccccc}
& \multicolumn{2}{c}{LC (10$^{15}$\,cm$^{-3}$)} & \multicolumn{2}{c}{MC (10$^{17}$\,cm$^{-3}$)}
& \multicolumn{2}{c}{HC (10$^{19}$\,cm$^{-3}$)} \\
& Ga-rich & N-rich & Ga-rich & N-rich & Ga-rich & N-rich \\ \hline
$E_F^{\mathrm{eq}}$ & 1.920 & 1.920 & 1.921 & 1.921 & 1.921 & 1.922 \\
$\alpha$ & 2.50$\times10^{-1}$ & 7.85$\times10^{3}$ & 2.48$\times10^{1}$ &
7.78$\times10^5$ & 2.48$\times10^{3}$ & 7.72$\times10^7$ \\
$\left[n\right]$ & 3.09$\times10^{13}$ & 3.09$\times10^{13}$ & 3.11$\times10^{13}$ &
3.11$\times10^{13}$ & 3.11$\times10^{13}$ & 3.14$\times10^{13}$ \\
$\left[p\right]$ & 8.07$\times10^{12}$ & 8.07$\times10^{12}$ & 8.00$\times10^{12}$ &
8.00$\times10^{12}$ & 8.00$\times10^{12}$ & 7.93$\times10^{12}$ \\
$\left[\mathrm{C_{Ga}^{1+}}\right]$ & $<10^{10}$ & 4.70$\times10^{12}$ & $<10^{10}$ &
4.62$\times10^{14}$ & $<10^{10}$ & 4.54$\times10^{16}$ \\
$\left[\mathrm{C_{I}^{1+}}\right]$ & $<10^{10}$ & $<10^{10}$ & $<10^{10}$ &
$<10^{10}$ & $<10^{10}$ & 6.54$\times10^{10}$ \\
$\left[\mathrm{V_N^{1+}}\right]$ & 1.00$\times10^{15}$ & 1.00$\times10^{15}$ & 1.00$\times10^{17}$ &
9.99$\times10^{16}$ & 1.00$\times10^{19}$ & 9.82$\times10^{18}$ \\
$\left[\mathrm{C_N^{1-}}\right]$ & \fbox{$\bm{1.00\times10^{15}}$} & \fbox{$\bm{9.95\times10^{14}}$} & \fbox{$\bm{1.00\times10^{17}}$} &
\fbox{$\bm{9.95\times10^{16}}$} & \fbox{$\bm{1.00\times10^{19}}$} & \fbox{$\bm{9.95\times10^{18}}$} \\
$\left[\mathrm{C_N^0}\right]$ & 1.21$
\times10^{11}$ & 1.20$\times10^{11}$ & 1.20$\times10^{13}$ &
1.19$\times10^{13}$ & 1.20$\times10^{15}$ & 1.18$\times10^{15}$ \\
$\left[\mathrm{C_N-C_{Ga}^0}\right]$ & $<10^{10}$ & $<10^{10}$ & $<10^{10}$ &
$<10^{10}$ & $<10^{10}$ & 7.59$\times10^{11}$ \\
$\left[\mathrm{V_{N}^0}\right]$ & $<10^{10}$ & 1.09$\times10^{10}$ & 1.08$\times10^{11}$ &
1.08$\times10^{12}$ & 1.08$\times10^{14}$ & 1.07$\times10^{14}$ \\
\end{tabular}
\end{ruledtabular}
\end{table*}

\begin{table*}
\caption{Concentrations of all defects (H, Si and O as well as C and vacancies)
in the system. Only the ones with more than
10$^{10}$\,cm$^{-3}$ are shown. Three different C concentrations are denoted as
LC (low C concentration), MC (medium C concentration) and HC (high C concentration).
$E_F^{\mathrm{eq}}$ denotes the equilibrium Fermi energy in eV. Highest C concentrations
in each condition are denoted in bold. All the concentrations
are in units of cm$^{-3}$.\label{t:concHSO}}
\begin{ruledtabular}
\begin{tabular}{ccccccc}
& \multicolumn{2}{c}{LC (10$^{15}$\,cm$^{-3}$)} & \multicolumn{2}{c}{MC (10$^{17}$\,cm$^{-3}$)}
& \multicolumn{2}{c}{HC (10$^{19}$\,cm$^{-3}$)} \\
& Ga-rich & N-rich & Ga-rich & N-rich & Ga-rich & N-rich \\ \hline
$E_F^{\mathrm{eq}}$ & 2.548 & 3.012 & 2.806 & 3.189 & 2.962 & 3.333 \\
$\alpha$ & 4.59$\times10^{-6}$ & 4.88$\times10^{-1}$ & 4.41$\times10^{-4}$ &
1.13$\times10^1$ & 3.93$\times10^{-2}$ & 3.25$\times10^2$ \\
$\left[n\right]$ & 7.62$\times10^{15}$ & 4.46$\times10^{17}$ & 7.32$\times10^{16}$ &
2.11$\times10^{18}$ & 2.88$\times10^{17}$ & 7.45$\times10^{18}$ \\
$\left[p\right]$ & 3.27$\times10^{10}$ & $<10^{10}$ & $<10^{10}$ &
$<10^{10}$ & $<10^{10}$ & $<10^{10}$ \\
$\left[\mathrm{H_I^{1+}}\right]$ & 2.68$\times10^{10}$ & 4.87$\times10^{13}$ & 2.68$\times10^{11}$ &
2.38$\times10^{14}$ & 6.09$\times10^{12}$ & 1.94$\times10^{15}$ \\
$\left[\mathrm{Si_{Ga}^{1+}}\right]$ & 6.18$\times10^{15}$ & 3.55$\times10^{17}$ & 6.19$\times10^{16}$ &
1.74$\times10^{18}$ & 1.40$\times10^{18}$ & 1.42$\times10^{19}$ \\
$\left[\mathrm{O_N^{1+}}\right]$ & 1.50$\times10^{15}$ & 9.70$\times10^{16}$ & 1.50$\times10^{16}$ &
4.75$\times10^{17}$ & 3.40$\times10^{17}$ & 3.86$\times10^{18}$ \\
$\left[\mathrm{C_N^{1-}}\right]$ & 4.52$\times10^{12}$ & \fbox{$\bm{8.93\times10^{14}}$} & 4.19$\times10^{15}$ &
\fbox{$\bm{9.75\times10^{16}}$} & 1.46$\times10^{18}$ & \fbox{$\bm{9.93\times10^{18}}$} \\
$\left[\mathrm{H_I^{1-}}\right]$ & $<10^{10}$ & 6.65$\times10^{13}$ & $<10^{10}$ &
7.26$\times10^{15}$ & 3.46$\times10^{12}$ & 7.39$\times10^{17}$ \\
$\left[\mathrm{V_{Ga}^{2-}}\right]$ & $<10^{10}$ & $<10^{10}$ & $<10^{10}$ &
$<10^{10}$ & $<10^{10}$ & 4.68$\times10^{10}$ \\
$\left[\mathrm{V_{Ga}^{3-}}\right]$ & $<10^{10}$ & $<10^{10}$ & $<10^{10}$ &
1.41$\times10^{11}$ & $<10^{10}$ & 1.79$\times10^{14}$ \\
$\left[\mathrm{C_N^0}\right]$ & $<10^{10}$ & $<10^{10}$ & $<10^{10}$ &
$<10^{10}$ & 1.90$\times10^{10}$ & $<10^{10}$ \\
$\left[\mathrm{C_N-H_I^0}\right]$ & 4.21$\times10^{11}$ & 1.42$\times10^{11}$ & 4.05$\times10^{12}$ &
3.28$\times10^{12}$ & 3.61$\times10^{14}$ & 9.45$\times10^{13}$ \\
$\left[\mathrm{C_{Ga}-H_I^0}\right]$ & $<10^{10}$ & $<10^{10}$ & $<10^{10}$ &
$<10^{10}$ & $<10^{10}$ & 2.93$\times10^{10}$ \\
$\left[\mathrm{Si_{Ga}-C_N^0}\right]$ & \fbox{$\bm{9.78\times10^{14}}$} & 1.04$\times10^{14}$ & \fbox{$\bm{9.42\times10^{16}}$} &
2.41$\times10^{15}$ & \fbox{$\bm{8.39\times10^{18}}$} & 6.95$\times10^{16}$ \\
$\left[\mathrm{O_N-C_N^0}\right]$ & 1.72$\times10^{13}$ & 2.06$\times10^{12}$ & 1.65$\times10^{15}$ &
4.77$\times10^{13}$ & 1.47$\times10^{17}$ & 1.37$\times10^{15}$ \\
\end{tabular}
\end{ruledtabular}
\end{table*}

In the case (a), $E_F^{\mathrm{eq}}$ is pinned at near the mid gap
(1.92\,eV) regardless of the C concentrations and growth conditions.
Among the donors, $\mathrm{V_N^{1+}}$ is the dominant one. In
addition, as the C concentration is increased,
$\mathrm{C_{Ga}^{1+}}$ and $\mathrm{C_I^{1+}}$ start to have
substantial presence with significant concentrations.
$\mathrm{C_N^{1-}}$ is the only acceptor-like defect with a
substantial amount present. In addition to these donor and acceptor
defects, $\mathrm{C_N^0}$, $\mathrm{V_N^0}$ and
$\mathrm{C_N-C_{Ga}^0}$ also exist as neutral defects. The position
of $E_F^{\mathrm{eq}}$ corresponds to the intersection of the
formation energies of the $\mathrm{C_N^{1-}}$ and
$\mathrm{V_N^{1+}}$ as shown in Fig.~\ref{f:EfdomCV}. The low
carrier concentrations resulting in his case leads to a
semi-insulating behavior of the system.

Next we consider the case (b). Among the donor defects,
$\mathrm{Si_{Ga}}^{1+}$ is the dominant impurity. This can be
expected from its low formation energy. Furthermore, this is the
dominant form of defect considered here regardless of the total
concentration of C and of the growth environment. Besides
$\mathrm{Si_{Ga}}^{1+}$, substantial amounts of $\mathrm{H_I^{1+}}$
and $\mathrm{O_N^{1+}}$ exist as donors. There are little C-related
donor defects unlike the case (a).
Among the acceptor defects, $\mathrm{C_N^{1-}}$ has the majority
presence. There are sizable amounts of $\mathrm{H_I^{1-}}$ in
particular in N-rich conditions, and $\mathrm{V_{Ga}^{2-}}$ and
$\mathrm{V_{Ga}^{3-}}$ increases when the $E_F^{\mathrm{eq}}$
approaches the CBM in N-rich conditions. Except for $\mathrm{C_N^0}$
with HC in Ga-rich conditions, neutral defects consist of C-related
complexes. Among them, $\mathrm{Si_{Ga}-C_N^0}$ is the dominant
form.

Focusing on the C-related defects, regardless of the total carbon
concentrations, $\mathrm{Si_{Ga}-C_N^0}$ is the dominant form in
Ga-rich conditions, whereas $\mathrm{C_N^{1-}}$ is the dominant form
in N-rich conditions. This means that when the total C concentration
increases, those of $\mathrm{Si_{Ga}-C_N^0}$ and $\mathrm{C_N^{1-}}$
increase as well. While the concentration of
$\mathrm{Si_{Ga}-C_N^0}$ has no dependency on the value of $E_F$,
the increase of $\mathrm{C_N^{1-}}$ results from the lower value of
its formation energy, leading to the higher value of the $E_F$. As a
result, the value $E_F^{\mathrm{eq}}$ becomes higher, when the total
C concentration is changed from low (LC) to high (HC).
Additionally, when the same C concentration is present both in
Ga-rich and N-rich conditions, $E_F$ is higher in N-rich conditions
than in Ga-rich conditions in order to have the same formation
energies in both situations. This is the reason why
$E_F^{\mathrm{eq}}$ is higher in N-rich conditions than in Ga-rich
conditions.
Comparing the two different cases, unintentional impurities such as
H, Si and O are the mostly responsible for $n$-type behavior in GaN
rather than native defects. If the amount of unintentional
impurities, in particular Si and O, is kept low, then the situation is
close to the case (a) and semi-insulating GaN is readily obtained.
In light of all these results, it is important to consider again the
experimental findings of Tompkins and coworkers.\cite{tompkins11} In
Ref.~\onlinecite{tompkins11}, the authors posit the presence of an
unknown forms of carbon in their GaN sample in addition to acceptor
type $\mathrm{C_N^{1-}}$. Our results indicate that the neutral
$\mathrm{Si_{Ga}-C_N}$ complex is a likely candidate for this
unknown form of carbon in GaN.

\section{Conclusion\label{s:conclusion}}

Various carbon based complexes in GaN were studied by using HSE
hybrid density functionals within the framework of generalized
Kohn-Sham density functional theory. Extending our previous work on
carbon--carbon/carbon-vacancy complexes, we have considered
complexes of carbon with other types of unintentional impurities,
i.e.\ hydrogen (H--C), silicon (Si--C) and oxygen (O--C)\@. The
structures of these complexes were fully optimized. From the
computed total energies, defect formation energies were obtained.
Using the formation energies, binding energies of complexes and the
transition levels were evaluated. Both thermal and optical
activation energies were derived from the transition levels to
compare our calculated results with the experimentally observed
carbon related trap levels in GaN\@.

Two types of H--C complexes were considered: $\mathrm{C_N-H_I}$ and
$\mathrm{C_{Ga}-H_I}$. The (+/0) transition level of $\mathrm{C_N-H_I}$ at 0.09\,eV above
the VBM has $E_{\mathrm{TH}}$=3.36\,eV and this may correspond to $E_c - 3.20$ level
observed by Shah \emph{et al.}~\cite{shah12}. The optical transition levels
(2+/+) and (+/0) of $\mathrm{C_{Ga}-H_I}$ has $E_{\mathrm{OPT}}$=2.13 and 2.56,eV, respectively.
The former corresponds to the experimentally observed $E_c-1.94/2.05$\,eV trap
level~\cite{armstrong05},
whereas the latter corresponds to the $E_c-2.7/2.8$\,eV trap level~\cite{polyakov13}.

Four different Si--C complexes were examined: $\mathrm{Si_{Ga}-C_N}$,
$\mathrm{Si_N-C_{Ga}}$, $\mathrm{Si_I-C_N}$ and $\mathrm{Si_{Ga}-C_I}$.
Among them, $\mathrm{Si_{Ga}-C_N}$ has the lowest formation energy.
Its (+/0) donor level at 0.27\,eV ($E_{\mathrm{TH}}$=3.18\,eV) corresponds
to the experimentally observed $E_c-3.20$\,eV trap level.
The (3+/2+) level of $\mathrm{Si_I-C_N}$ has 2.65\,eV transition level,
which can be assigned to $E_c-2.7/2.8$\,eV level~\cite{polyakov13}. Finally,
the (+/0) of $\mathrm{Si_{Ga}-C_I}$ level has $E_{\mathrm{OPT}}$=1.43\,eV optical
activation energy and this value is very close to $E_c-1.35$\,eV and $E_c-1.3/1.4$\,eV
levels observed by Armstrong \emph{et al.}~\cite{armstrong05} and by Polyakov \emph{et al.}~\cite{polyakov13}.
Note that $\mathrm{Si_N-C_{Ga}}$ complex was excluded from the assignment because it has low concentration with high formation energy compared to other Si--C complexes.

Five different O--C complexes were studied. $\mathrm{O_I-C_N}$ and $\mathrm{O_N-C_I}$
has the same configuration after the full geometry optimization. Thus, four different
complexes were considered in practice.
%
The (2+/+) level of $\mathrm{O_I-C_N/O_N-C_I}$ is located at 0.75\,eV ($E_{\mathrm{TH}}$=2.70\,eV) above the VBM.
Experimentally observed $E_v+0.9$\,eV~\cite{armstrong05} and $E_v+0.86$\,eV~\cite{honda12}
levels as well as $E_c-2.69$\,eV can be assigned to this level. 
The (0/$-$) level of $\mathrm{O_I-C_N/O_N-C_I}$ has $E_{\mathrm{TH}}$=0.57\,eV and
$E_{\mathrm{OPT}}$=1.53\,eV. The former can be assigned to the thermally detected
$E_c-0.40$\,eV level~\cite{honda12} and the latter has close energy to the optically observed
$E_c-1.35$\,eV and $E_c-1.3/1.4$\,eV levels.
Finally the (+/0) level of the $\mathrm{O_N-C_N}$ complex, which is a second nearest
neighbor pair, has $E_{\mathrm{OPT}}$=3.40\,eV, which can be assigned to the $E_c-3.28$\,eV
trap level.
As in the case of $\mathrm{Si_N-C_{Ga}}$, $\mathrm{O_{Ga}-C_N}$ complex was excluded from the assignment because it has low concentration with high formation energy compared to other O--C complexes.

From the analysis of the impurity concentration we have found that
when only carbon and vacancies are considered as electrically active
defects, Fermi energy is pinned near the mid gap with 1.92\,eV value
regardless of the total carbon concentrations and growth conditions.
This equilibrium Fermi energy corresponds to the intersection of the
formation energies of $\mathrm{V_N^{1+}}$ and $\mathrm{C_N^{1-}}$.
In this situation carrier concentration is low and semi-insulating
behavior can be expected. Other than the $\mathrm{C_N}$ and
$\mathrm{V_N}$, $\mathrm{C_{Ga}}$, $\mathrm{C_I}$ and
$\mathrm{C_N-C_{Ga}}$ show substantial presence with increasing
total carbon concentrations.

When hydrogen, silicon and oxygen are also considered as electrically active defects
besides carbon and vacancies, The situation is drastically changed.
The equilibrium Fermi energy is closer to conduction band. It is $\sim$2.5\,eV with
low carbon concentration situation in Ga-rich conditions and it becomes $\sim$3.3\,eV
with high carbon concentration situation in N-rich conditions. The carrier concentration
is much higher, as high as 7.45$\times10^{18}$\,cm$^{-3}$. Main donor defects are H, Si
and O, in particular $\mathrm{Si_{Ga}}$. No C-related donors have substantial concentrations.

As for acceptors, $\mathrm{C_N^{1-}}$ is dominant. $\mathrm{H_I}$
and $\mathrm{V_{Ga}}$ also have substantial concentrations with
increasing carbon concentrations. Neutral defects are mostly
C-related complexes, among which $\mathrm{Si_{Ga}-C_N^0}$ is the
most dominant form. This neutral complex is a likely candidate for
the unknown form of carbon in GaN observed in experimental results.

\begin{acknowledgments}



The authors thank K. Jones and R. Tompkins of the Army
Research Laboratory, T. Moustakas of Boston University and R. Kaplar
of Sandia National Laboratory for many discussions and their help
in understanding the experimental techniques. The authors are grateful
to A. Wright, S. Lee and N. Modine from Sandia National Laboratory
for discussing the result of our work.
The authors gratefully acknowledge financial support from the U. S.
Army Research Laboratory through the Collaborative Research Alliance
(CRA) Grant No.\ W911NF-12-2-0023 for MultiScale multidisciplinary Modeling of Electronic Materials
(MSME). This work was performed using DoD HPCMP supercomputing resources
and computational resources provided by the 2014 Army Research Office
Grant No.\ W911NF-14-1-0432 DURIP Award made
to E. Bellotti.

\end{acknowledgments}


\begin{thebibliography}{42}
\bibitem [{FDm()}]{FDmethod}%
  \BibitemOpen
  \href@noop {} {}\bibinfo {note} {Finite difference method implemented in the
  VASP code is used. Both positive and negative displacements by 0.015\,\AA\
  are introduced in the calculations.}\BibitemShut {Stop}%
\end{thebibliography}
\end{document}